\documentstyle[epsf,11pt,fleqn]{article}
\newcommand{\befig}{ \begin{figure}[htbp] }
\input epsf

\begin{document}
\renewcommand{\topfraction}{0.9}
\renewcommand{\bottomfraction}{0.9}
\renewcommand{\textfraction}{0.1}

\title{How Chaotic is the Stadium Billiard? \\ A Semiclassical Analysis}
\author{
Gregor Tanner\thanks{email: tanner@kaos.nbi.dk}\\
Niels Bohr Institute, Blegdamsvej 17, \\
DK-2100 Copenhagen \O , Denmark\\
}
\date{September 26, 1996}
\maketitle

\begin{abstract}

The impression gained from the literature published to date
is that the spectrum of the stadium billiard can be adequately described,
semiclassically,
by the Gutzwiller periodic orbit trace formula together with a
modified treatment of the margin\-al\-ly stable family of 
bouncing ball orbits.
I show that this belief is erroneous.
The Gutzwiller trace formula is not applicable for the phase space dynamics 
\underline{near} the bouncing ball orbits.
Unstable periodic orbits close to the marginally stable family in phase 
space cannot be treated as isolated stationary phase points when
approximating the trace of the Green function.
Semiclassical contributions to the trace  
show an $\hbar$ -- dependent transition from hard chaos to integrable 
behavior for trajectories approaching the bouncing ball orbits.
A whole region in phase space surrounding the marginal stable family
acts, semiclassically, like a stable island with boundaries
being explicitly $\hbar$--dependent.
The localized bouncing ball states found in the billiard 
derive from this semiclassically stable island. 
The bouncing ball orbits themselves, however, do not
contribute to individual eigenvalues in the spectrum. 
An EBK--like quantization of the regular bouncing ball eigenstates in the 
stadium can be derived. 
The stadium billiard is thus an ideal model for studying the influence
of almost regular dynamics near marginally stable boundaries on 
quantum mechanics. 
This behavior is generically found 
at the border of classically stable islands in systems with a mixed phase 
space structure. \\

\end{abstract}


\section{Introduction} \label{intro}

The derivation of semiclassical periodic orbit formulas for the trace
of the quantum Green function led to a deeper understanding of the influence
of classical dynamics on quantum spectra. Closed periodic orbit expressions
have been given by Gutzwiller \cite{Gut90} and Balian and Bloch \cite{Bal72} 
for ``hard chaos'' systems and by Berry and Tabor \cite{Ber76,Ber77} for 
integrable
dynamics. Integrability and hard chaos represent the two extremes on the
scale of possible Hamiltonian dynamics. The term ``hard chaos'' introduced 
by Gutzwiller \cite{Gut90} is, however, not well defined. It implies, that 
all periodic orbits are unstable and ``sufficiently'' isolated to allow for 
the stationary phase approximations in the derivation of the trace formula. 
Hence, for lack of a better definition, one might say that a system exhibits 
``hard chaos'' if the Gutzwiller
trace formula as it stands is the leading term in a semiclassical expansion
in $\hbar$. 

Neither integrability nor hard chaos is generic in low dimensional bounded
Hamiltonian dynamics.  The variety of classical systems proposed as testing 
models for the Gutzwiller trace formula in the last decade exemplify the 
problem of finding ideal chaos in the sense described
above. Typical Hamiltonian systems show a mixture of the two extremes. 
Chaotic regions in phase space (containing unstable periodic orbits only) are 
interspersed by stable islands, which themselves may have a complicated inner 
structure consisting of chaotic bands and invariant tori. The 
``stable regions'' reflect locally almost integrable 
behavior  in phase space and can be treated semiclassically by an 
Einstein--Brillouin--Keller (EBK) approximation \cite{Gut90,Boh93,Win94}. 
Semiclassical contributions to the trace of the Green function from 
periodic orbits inside the stable islands are of modified Berry--Tabor--type 
as has been discussed in \cite{Ozo87,Tom95,Ric96,Sie96,Cre96}. 

Much less is known about a semiclassical treatment of the transition region 
from the stable island to the outer chaotic neighborhood. The outer boundary 
acts as a classically impenetrable wall in two degrees of freedom. A coupling 
between dynamically separated regions can only be described by complex 
trajectories \cite{Bal72,Shu95,Dor95}. Although the outer 
chaotic region contains (per definition) only unstable periodic orbits, it
shows almost regular behavior near the stable components. The characteristics
for this kind of regularity, also called intermittency in the chaos 
literature \cite{Sch89}, is a vanishing Liapunov exponent $\lambda_p$ 
for unstable periodic orbits approaching the island, i.e.\
\[ \lambda_p = \frac{\log\Lambda_p}{T_p} \; \longrightarrow
\!\!\!\!\!\!\! \!\!\!\!\!\!
{~\atop {\scriptstyle T_p\to\infty}} \; 0 ,\]
where $\Lambda_p$ denotes the largest eigenvalue of the stability matrix 
(describing the linearized dynamics in the neighborhood of the periodic 
orbit) and $T_p$ is the period.

I will show that intermittency in bounded systems leads to leading order 
corrections in the Gutzwiller trace formula. Semiclassical contributions 
from (unstable) periodic orbits approaching  the boundary of a stable island 
show a $\hbar$--dependent
transition from Gutzwiller to Berry -- Tabor like behavior. This allows one
to extend the concept of EBK--quantization from a stable region over 
the marginal stable boundary into its ``chaotic'' neighborhood.\\

We will derive these results for a specific example, the quarter stadium
billiard, (see Fig.\ \ref{fig:stadium}). This might appear surprising on 
first sight, because this billiard has been introduced as an example of a 
dynamical system which is completely ergodic with positive Liapunov exponent 
\cite{Bun79}.  The stadium billiard has thus been regarded as an ideal model 
for a ``chaotic'' system. First indications for the validity of the random 
matrix conjecture relating the level statistics of individual ``chaotic'' 
systems to those of an ensemble of random Gaussian orthogonal matrices have 
been found in this billiard \cite{McD79,Cas80, Boh84}. Also the discovery of 
scarred wave functions \cite{Hel84} 
was first made in this system. From then on, the classical and
quantum aspects of the stadium billiard were studied intensively both 
theoretically and in microwave experiments \cite{Sto90,Gra92}, making 
it a standard model in quantum chaology. The billiard dynamics possesses, 
however, regularities, which prevents it from being a hard chaos system. 
The so--called bouncing ball orbits, i.e.\ the continuous family of periodic 
orbits running back and forth in the rectangle, are marginally 
stable. Another peculiarity of the stadium is the existence of 
whispering gallery orbits accumulating at the boundary of the billiard. These
orbits have a strong, non--generic influence on the spectral statistics which 
causes deviations from the GOE--predictions \cite{Gra92}. There also exist 
quantum eigenstates, which are localized along the bouncing ball orbits 
(the bouncing ball states), or along the boundary (the whispering gallery 
states), in a much more pronounced way than the scars found
along other periodic orbits.

As a consequence, the stadium billiard is not an example of a hard
chaos billiard, but may serve as an ideal model for systems exhibiting both 
regular and chaotic dynamics. Though there is no stable island, the bouncing 
ball orbits act as the boundary of a torus enclosing an area with zero volume 
in phase space. This allows us to study pure boundary effects without dealing 
with the inner structure of the island. 

Semiclassical results obtained for hard chaos systems suffer considerable 
changes when dealing with intermittency as will be shown here. First of all,
the contributions of the marginal stable family to the trace of the 
Green function need a special treatment \cite{Sie93,Alo94}. This is true,
however, also for unstable periodic orbits close to the bouncing ball family.
They can no longer be treated as isolated stationary phase points when 
approximating the trace of the Green function at finite $\hbar$. As a 
consequence, the Gutzwiller periodic orbit formula is not valid for the 
dynamics in a whole phase space volume surrounding the marginal stable family.
The size of this volume is explicitly $\hbar$ dependent and shrinks to 
zero in the semiclassical limit. 

Furthermore, standard semiclassical arguments referring to a cut--off in 
periodic orbit sums at a period given by the Heisenberg time 
are not valid for intermittent systems. Intermittency is related to a loss
of internal time scales in the dynamics and infinitely long orbits thus 
contribute dominantly to the semiclassical zeta function at finite $\hbar$. 
In particular, the regular bouncing ball states follow scaling laws for a 
cut--off in the orbit summation different from the estimate given
by the Berry -- Keating resummation technique \cite{Ber90}.\\
 
The article is organized as follows: in section \ref{sec:bb}, the 
contribution of the bouncing ball family to the trace of the Green function 
is discussed. The derivation follows mainly the ideas developed in 
\cite{Sie93,Alo94} although here we do not deal with diffraction effects. It 
will be 
shown that the bouncing ball orbits give no contribution to individual 
eigenvalues of the stadium billiard. In section \ref{sec:T-oper}, the 
Bogomolny transfer operator method \cite{Bog92} is introduced. 
A Poincar\'e surface of section is defined, which allows us to study the 
near  bouncing ball dynamics, but explicitly excludes the bouncing ball
family. The  Poincar\'e map in the bouncing ball limit is derived, 
which allows us to obtain the leading contributions to the transfer operator
for the bouncing ball spectrum. An approximate EBK--quantization of the 
bouncing ball states can be derived. 
An analysis of the trace of the transfer operator unveils the breakdown of
the Gutzwiller periodic orbit formula for unstable periodic orbits 
close to the bouncing ball family. 
In section \ref{sec:bbstates}, we develop the concept of a ``semiclassical 
island of stability'' surrounding the marginally stable bouncing ball family 
in phase space and discuss the semiclassical limit of the bouncing ball
spectrum.

\section{The bouncing ball orbits} \label{sec:bb}
In this section, the bouncing ball contribution to the quantum spectral
determinant
\begin{equation} \label{specdet}
D(E) = \exp\int_0^E dE'\mbox{Tr} G(E') = \prod_n (E - E_n) 
\end{equation}
is derived. The \{$E_n$\} denote the real quantum eigenvalues of the system.
Following Ref.\ \cite{Sie93,Alo94}, the Green function for the quarter
stadium billiard, Fig.\ \ref{fig:stadium},  
is divided into a bouncing ball part and the rest,  
\begin{equation} \label{Green}
G(q,q',k) = G_{bb}(q,q',k) + G_r(q,q',k), \end{equation}
where $k = \sqrt{2 m E}/\hbar$.
The bouncing ball contribution is restricted to $q$, $q'$ inside the rectangle
and contains in semiclassical approximation all paths from $q$ to $q'$ 
not reaching the circle boundary or bouncing off the vertical line in the 
rectangle. The classical paths correspond essentially to free motion,
and the bouncing ball Green function $G_{bb}$ can be written as an infinite
sum over Hankel functions (plus phases due to hard wall 
reflections). 
The trace then has the form \cite{Sie93}
\begin{equation} \mbox{Tr} G_{bb}(k) = g_{bb}(k) = -i\frac{a b}{2} k + 
i \frac{a}{2} - i a b k
\sum_{n=1}^{\infty} H_0^{(1)}(2 b k n) + \lim_{\epsilon \to 0} 
\frac{a b k}{2} N_0(\epsilon k), 
\label{trg}
\end{equation}
where $a$ is the length of the rectangle, $b$ denotes the radius of the 
circle and $H_0^{(1)}$ is the Hankel function of the first kind. The real part
of the trace diverges due to the logarithmic singularity of the Neumann 
function $N_0$ at the origin. This kind of divergence is a well known 
phenomenon for the trace of Green functions and can be regularised e.g.\ by 
defining 
$\tilde{g}_{bb}(E) = g_{bb}(E) - {\rm Re}[g_{bb}(E_0)]$ for $E_0$ fixed
\cite{Vor87,Kea94}. 
Note, that the same regularisation techniques have to be applied for the true
 quantum trace $g(E) = \sum_n (E-E_n)^{-1}$ and are thus not an artifact of
the semiclassical approximation. We now move directly to the integrated trace,
which  reads 
\begin{eqnarray} \label{inttr}
Ig_{bb}(k) = \int_0^k dk' \tilde{g}_{bb}(k') &=& B_{bb}(k^2) - 
	      i \pi \overline{N}_{bb}(k)
      - i \frac{a k}{2} \sum_{n=1}^{\infty} \frac{1}{n} H_1^{(1)}(2 k b n).\\
\label{inttrap}
                &\approx& B_{bb}(k^2) - i \pi \overline{N}_{bb}(k)
          - i \frac{a}{2}\sqrt{\frac{k}{\pi b}} \sum_{n=1}^{\infty} 
              \frac{1}{n^{3/2}} e^{2 i k b n-\frac{3}{4} i \pi},
\end{eqnarray}
and the asymptotic form of the Hankel function has been used in the last 
step. The sum in (\ref{inttrap}) is of the same kind as the contribution of 
a periodic orbit family in an integrable system (here the rectangle) 
as derived by Berry and Tabor \cite{Ber76,Kea87}. The real part of the 
non--oscillating contribution is a function of $k^2$ only, 
\begin{equation} \label{B2}
B_{bb}(k^2) = 
\frac{a b}{4\pi} k^2 (\log k^2 - 1) + 
          \frac{a b}{2}\beta k^2  + \frac{a}{b}\frac{\pi}{12}.
\end{equation}
The real parameter $\beta$ originates from the regularisation and has the form
\begin{equation}
\beta = -\frac{\log k_0}{\pi} - \sum_{n=1}^{\infty} N_0^{(1)}(2 b k_0 n)
\qquad \mbox{with} \quad k_0 = \sqrt{2 m E_0}\hbar.
\end{equation}
The imaginary part is the bouncing ball contribution to the mean level 
staircase function 
\begin{equation}\label{Nmean}
\overline{N}_{bb}(k) = \frac{a b}{4\pi} k^2 - \frac{2 a}{4\pi} k .
\end{equation}
The term $B_{bb}(k^2)$ is of the general form as derived in \cite{Kea94}
for arbitrary billiards with compact domain, i.e.\ 
\begin{equation} \label{B2g}
B(k^2) = \frac{A}{4 \pi} k^2(\log k^2 - 1) + \beta k^2 + \gamma \log k^2 +
c ,
\end{equation}
where $A$ corresponds to the volume of the billiard and  $\beta$, $\gamma$, 
$c$ are real constants. Note, that the logarithmic singularity for $k\to 0$
in (\ref{B2g}) is absent in (\ref{B2}). The sum in (\ref{inttr}) is convergent
on the whole complex $k$--plane (after analytic continuation for $Im(k) < 0$).
The function $Ig_{bb}(k)$ is free of singularities and analytic everywhere 
apart from the lines $Re(k) = m \pi/b$, $m$ integer, where it exhibits 
square -- root cusps. The dominant oscillation introduced through the 
bouncing ball orbits is clearly visible in the oscillatory part of the level
staircase function \cite{Gra92,Alo94}, and agrees with the prediction 
(\ref{inttr}). The bouncing ball part does, however, not contribute
to individual eigenvalues.
\begin{figure}
         \epsfxsize=14cm
         \epsfbox{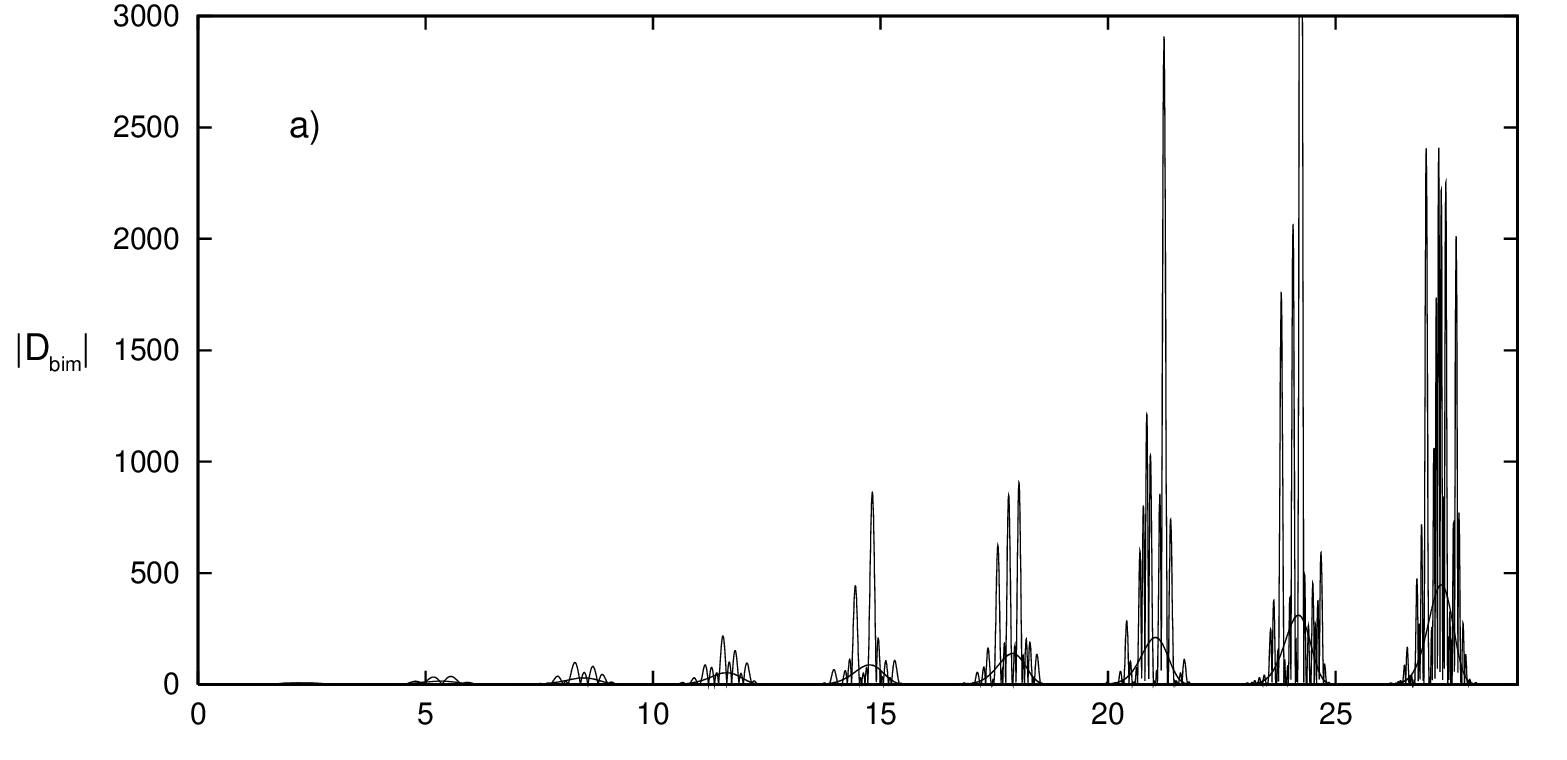}
         \epsfxsize=14cm
         \epsfbox{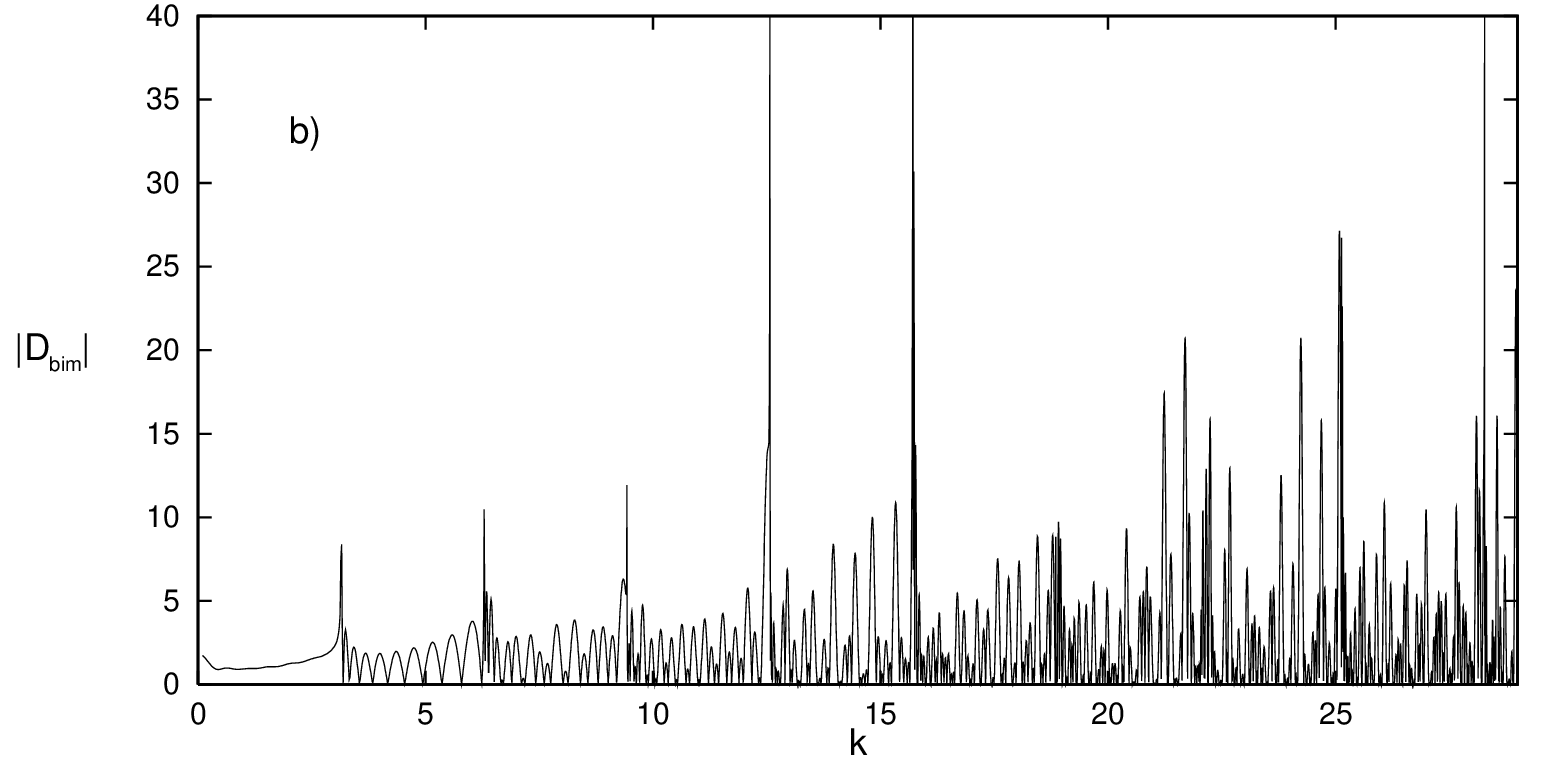}
         \caption[]
            {The modulus of the boundary integral spectral determinant 
	     before (a) and after (b) dividing out the bouncing ball part, 
	     Eqn.\ (\ref{specfak}), here for $a=5$, $b=1$; (in addition
	     $D_{bb}(k)$ is plotted in (a)). The $k$ -- interval shown
	     contains about 370 energy levels, which are the real zeros of the
	     spectral determinant.
	  }
\label{fig:det}
\end{figure}
The spectral determinant (\ref{specdet}) can be factorised according to
\begin{equation} \label{specfak}
D(k) \approx D_{bb}(k) D_r(k) \qquad \mbox{with} \quad 
D_{bb}(k) = e^{Ig_{bb}(k)},
\end{equation}
and the bouncing ball contribution $D_{bb}(k)$ is a function which has no 
zeros in the whole complex $k$ plane.
It gives rise to oscillations in the modulus of the spectral determinant with 
an amplitude growing like 
$\exp\left(\frac{a}{2}\sqrt{\frac{k}{\pi b}}\right)$ 
and a period $\pi/b$, but it does not influence the individual zeros.
The bouncing ball contributions to the spectral determinant can thus be 
divided out without losing information about the spectrum. This can be 
demonstrated by considering the boundary integral determinant
\begin{equation}\label{detbim}
D_{bim}(k) = \det({\bf 1} - {\bf K}(q,q',k)),
\end{equation}
where $\bf K$ is the boundary integral kernel defined for $q$ and $q'$ on
the boundary of the billiard, (for details see \cite{Boa94}). The function
$D_{bim}(k)$ has the same zeros as the exact spectral determinant $D(k)$, 
but may differ otherwise. Especially, a smooth part 
$\exp(B(k^2) - i\pi\overline{N}(k))$ is absent in the boundary integral 
determinant. This term originates from the zero length limit in the
Green function and is thus a pure volume effect. The modulation
due to the bouncing ball contribution is clearly
present in $D_{bim}(k)$, see Fig.\ \ref{fig:det}a. The large overall 
oscillation of the amplitude covers 7 order of magnitudes for $k\approx 30$ 
but is completely removed by factorising out the bouncing ball part 
$D_{bb}(k)$, see fig.\ \ref{fig:det}b. Note, that the boundary 
integral determinant after factorization exhibits cusps at $k = m \pi/b$, and 
is thus non--analytic there. \\

As a somewhat paradoxical result, we obtain that the bouncing ball family is 
responsible for strong modulations in the trace as well as in the spectral 
determinant. It does not, however, contribute to individual eigenvalues, and 
does not explain the regular modulation in the level spacing itself. The 
bouncing ball family covers a volume of measure zero in phase space, and is 
not sufficient to form a support for an eigenfunction alone. (This is of 
course true for any invariant family e.g.\ in an integrable system.) As a 
consequence, the neighborhood of the bouncing ball family in phase space 
must be responsible 
for the existence of the bouncing ball states and for the periodic 
change in the level spacings leading to the modulation in the level staircase 
function. The next section is devoted to a 
proper semiclassical analysis of the near bouncing ball dynamics. 

\section{The semiclassical transfer operator} \label{sec:T-oper}
So far, it has been widely believed, that the remainder term in the 
trace of the Green function (\ref{Green}) as well as in the 
spectral determinant (\ref{specfak}) can in leading semiclassical
approximation be described by the Gutzwiller periodic orbit formula
\cite{Gut90,Bal72} or equivalently by the semiclassical spectral determinant
\cite{Vor88}. This belief is confirmed by studies of the Fourier
transform of the (smoothed) quantum spectrum, which agrees reasonably well
with periodic orbit predictions \cite{Gra92,Sie93,Alo94}. The semiclassical 
expression for the determinant, valid for systems with unstable, isolated 
periodic orbits only, has the form 
\begin{equation} \label{detGut}
D(E) = A(E) e^{-i\pi\overline{N}(E)} \exp\left(
- \sum_p\sum_{r=1}^{\infty} \frac{\exp(i r S_p(E)/\hbar - i r\alpha_p \pi/2)}
{r \sqrt{|\det({\bf 1} - {\bf M}^r(E)_p)|}}\right).
\end{equation}
The sum is taken here over all (single repeats of) periodic orbits of 
the system. The classical action $S_p =\oint p dq$ is taken along the 
periodic orbit, and $\bf M$ denotes the reduced Monodromy matrix, which 
describes the linearized dynamics in phase space perpendicular to the 
periodic orbit on the energy manifold. The Maslov index $\alpha$ counts twice
 the full rotations  of the (real) eigenvectors of $\bf M$ around the orbit 
(plus two times the number of hard wall reflections). The prefactors in front
of the periodic orbit product are due to the zero-length limit in the Green 
function in a similar way as derived in (\ref{B2}), (\ref{Nmean}) and do not
contribute to the spectrum. Eqn.\ (\ref{detGut}) is formal in the sense, that 
it is not convergent for real energies \cite{Eck89} and suitable resummation 
techniques have to be applied \cite{Cvi88,Ber90,Tan91,Sie91}. Thereby, the 
exponential function containing the periodic orbit sum is expanded and the 
resulting terms are regrouped by ordering them with respect to the total 
action or the total symbol length (after choosing a suitable symbolic 
dynamics). Such techniques have been shown to work successfully for hard 
chaos systems. 

In the following, I will show that formula (\ref{detGut}) is not valid for
periodic orbits within a phase space region surrounding the bouncing ball 
family and will give explicit bounds for this area.

\subsection{The Bogomolny Transfer Operator} \label{sec:Bog_Tran}
Our starting point is the Bogomolny transfer operator \cite{Bog92} which is a 
semiclassical propagator for a classical Poincar\'e map. It has the form
\begin{equation} \label{BoT}
T(q,q',E) = \frac{1}{(2\pi i \hbar)^{(f-1)/2}} \sum_{cl.tr q\to q'} 
\sqrt{\left|\frac{\partial^2 S}{\partial q \partial q'} \right|}
e^{i S(q,q',E)/\hbar - i \nu \pi/2},
\end{equation}
where $q$, $q'$ are points on an appropriate Poincar\'e surface of section 
in coordinate space and $f$ denotes the number of degree of freedom.
The sum has to be taken over all classical paths from $q$ to $q'$ crossing
the Poincar\'e surface only once with momentum pointing in the direction of
the normal to the surface. Again, $S(q,q';E)$ denotes the classical action  
along the path for fixed energy $E$. For billiard systems, it equals $k$ 
times the length of the classical trajectory. The integer number $\nu$ counts
the number of caustics in $q$--space (plus again twice the number of hard 
wall reflections). The semiclassical eigenvalues are given by the zeros of 
the determinant $\mbox{det}({\bf 1} - {\bf T}(q,q',E))$. Evaluating the 
determinant in a cumulant expansion using the Plemelj--Smithies formula 
\cite{Bog92,Wir95}
\begin{eqnarray} \label{cumexp}
\det({\bf 1}-{\bf T}) &=& 
           \sum_{n=0}^{\infty} (-1)^n \frac{\alpha_n({\bf T})}{m!}\\ \nonumber
	             &=& 1 - \mbox{Tr} {\bf T} - \frac{1}{2}
		     (\mbox{Tr}{\bf T}^2 - (\mbox{Tr}{\bf T})^2 ) - \ldots
\end{eqnarray}
with
\begin{equation}
\alpha_n({\bf T}) = \left| \begin{array}{ccccc}
                 \mbox{Tr} {\bf T}  & n-1              & 0 &\ldots&0\\
                 \mbox{Tr} {\bf T^2}&\mbox{Tr} {\bf T} &n-2&\ldots&0\\
  \mbox{Tr} {\bf T^3}&\mbox{Tr} {\bf T^2} &\mbox{Tr} {\bf T}&\ldots&0\\
	                  \vdots&\vdots&\vdots&\vdots&\vdots\\
  \mbox{Tr} {\bf T^{n}}&\mbox{Tr} {\bf T^{n-1}} &\mbox{Tr} {\bf T^{n-2}}&
  \ldots&\mbox{Tr} {\bf T}
   \end{array} \right|
\end{equation}
provides the connection to the expanded periodic orbit formula (\ref{detGut})
using the iterates of the map as an expansion parameter.
Periodic orbits appear as stationary phase points in the various traces and
the amplitudes are recovered using the relation 
\begin{equation} \label{detM}
\frac{1}{\sqrt{|\det({\bf 1} - {\bf M})|}} = 
\sqrt{\left|\frac{\partial^2 S(q,q')}{\partial q \partial q'} \right|_{q=q'}} 
\left/\sqrt {\left|\frac{\partial^2 S(q,q)}{\partial q^2 } \right|} \right. .
\end{equation}
The stationary phase approximation demands periodic orbits
to be unstable and sufficiently isolated. The last condition will
be discussed in detail later. Note that the determinant 
and the cumulant expansion (\ref{cumexp}) are well
defined only if the operator $\bf T$ is {\em trace class}, which means
essentially, that the trace of $\bf T$ exists and is finite in any basis.
(For further details see \cite{Wir95}, other expansion methods using Fredholm
theory are discussed in \cite{Geo95}).

\begin{figure}[tb]
\centerline{
         \epsfxsize=13.0cm
         \epsfbox{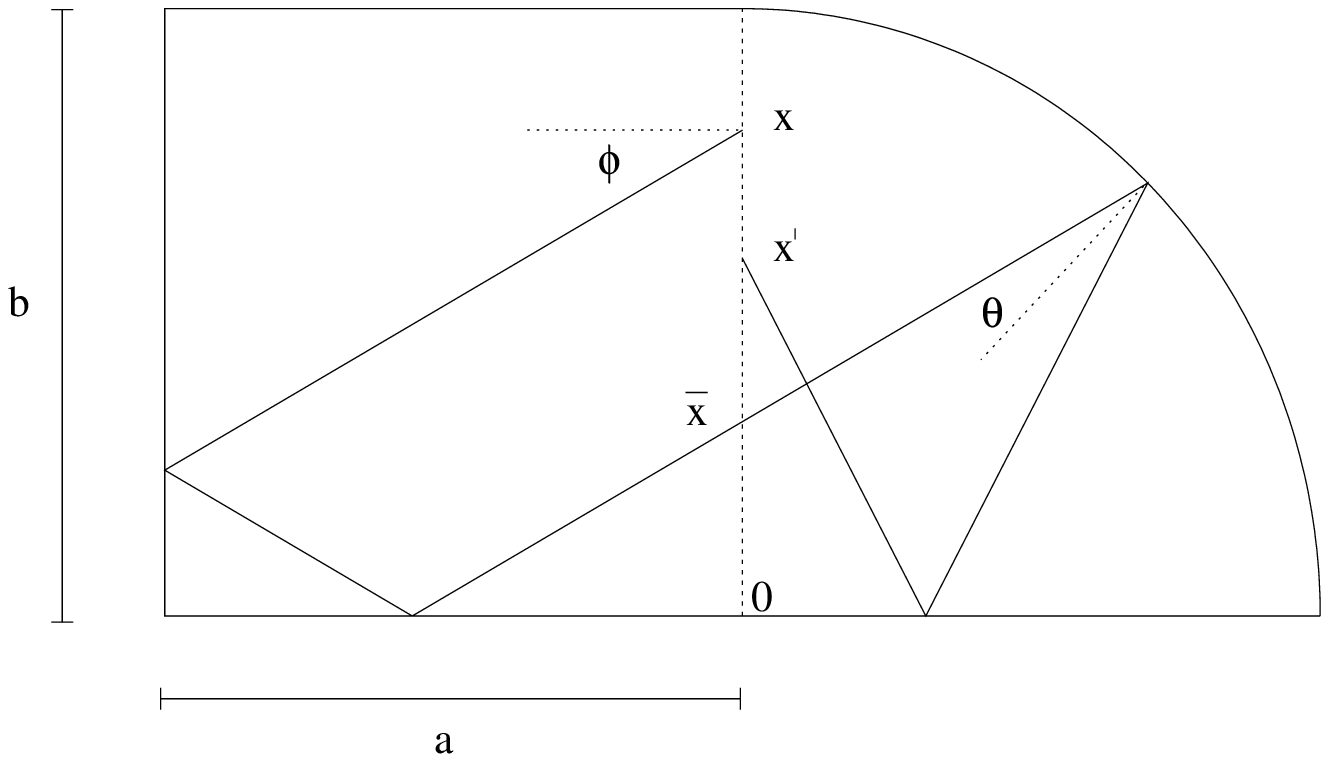}
         }
         \caption[]
            {
 	    The quarter stadium billiard; the trajectory shown corresponds
	    to ($m$,$l$) = ($-1$,1).
	  }
\label{fig:stadium}
\end{figure}

The Bogomolny transfer matrix method has been shown to work satisfactorily 
for hard chaos \cite{Sze94}, mixed \cite{Sze94,Hag95} as well
as integrable \cite{Lau92,Boa94,Sze94} systems, and has also been applied
successfully to the stadium taking the billiard boundary as the Poincar\'e 
surface of section \cite{Boa94}. We seek now a Poincar\'e map which 
reflects the whole dynamics but excludes the bouncing ball orbits in order 
to obtain a semiclassical expression for the remainder $D_r$ in 
(\ref{specfak}). Choosing the intersection between the rectangle and the 
circle
(the vertical dotted line in fig.\ \ref{fig:stadium}) fulfills this criteria.
The transfer operator corresponding to this Poincar\'e section takes on a 
particularly simple form for both the near bouncing ball limit and 
the whispering gallery limit. It is thus preferable to other possible choices
such as e.g.\ the circle boundary.

\subsection{The classical Poincar\'e Map} \label{sec:classPM}
In the following, I will restrict attention to the stadium with $b$ = 1. The 
spectrum for general $b$ is obtained by simple scaling relations.
I will consider the classical Poincar\'e map $(x,\phi) \to (x',\phi')$ with 
$x\in [0,1]$ being the coordinate on the Poincar\'e plane starting from the 
bottom line. The angle $\phi \in ]-\pi/2,\pi/2[$ corresponds to the momentum 
vector pointing away from the circle measured in the clockwise direction, 
see Fig.\ \ref{fig:stadium}. (The corresponding energy dependent area 
preserving map is obtained in the coordinates $(x,p_x)$ = 
$(x,\sqrt{2 m E} \sin\phi)$.) The map can be written as
\begin{eqnarray} \label{PM}
\overline{x} &=& (-1)^m \left[x+2 a \tan\phi - 
(m + \frac{1 - (-1)^m}{2})\right] \\ \nonumber
\overline{\phi} &=& (-1)^m \phi \quad \qquad \qquad \qquad \mbox{with} 
\quad m\le x+2 a \tan\phi < m+1 \\\nonumber
\theta &=& \arcsin(\overline{x}\cos\overline{\phi})\\ \nonumber
\phi' &=& \overline{\phi} + 2(l+1)\theta - k\pi \qquad \mbox{with}
\quad l\le\frac{\overline{\phi} + \theta}{\pi - 2 \theta} < l+1 \\ \nonumber
x' &=& \overline{x}\; \frac{\cos\overline{\phi}}{\cos\phi'}.
\end{eqnarray}
The coordinates $(\overline{x},\overline{\phi})$ correspond to the first 
return at the Poincar\'e plane with momentum pointing toward the circle.
The angle $\theta$ is the angle of incidence for reflections on the circular
section of the boundary; see Fig.\ \ref{fig:stadium}.

The length of a trajectory for one iteration of the map is
\begin{equation}
L(x,\phi) = \frac{2 a + \cos(\overline{\phi}+\theta)}{\cos\phi} + 
2 l \cos\theta + \frac{\cos(\phi'-\theta)}{\cos\phi'}.
\end{equation}
The general expression for the Monodromy matrix is given in Appendix A. 
The integer numbers $m \in [-\infty,\infty]$ correspond to $|m|$--reflections 
on the bottom or top line in the rectangle, the sign of $m$ equals the
sign of $\phi$. The index $l\in [0,\infty]$ counts the number of free flights 
in the circle and corresponds to $(l+1)$ -- bounces with the circle boundary.
The total number of bounces $n_{tot}$ with the billiard boundary is thus
\begin{equation} \label{nbounce}
n_{tot} = |m| + (l + 1) + 2 .
\end{equation}

The map provides automatically a symbolic coding. It should be noted that the
code does not form a ``good'' symbolic 
dynamics in the sense of Markov partition theory. Multiple iterates of the 
map are not uniquely encoded by symbol strings 
$\ldots,(m,l)_i,(m,l)_{i+1},\ldots$,
i.e.\ the partition is not generating.  In addition, there is strong pruning, 
i.e. some of the possible symbols strings are not realized by trajectories of 
the map. The symbols defined by the map (\ref{PM}) 
do, however, reflect the important contributions of the dynamics to a 
semiclassical description as will be seen in the next section. The problem of 
finding a ``good'' symbolic dynamics for the stadium \cite{Bih92,Han93} might 
be a consequence of this fundamental dilemma. 

\begin{figure}[tb]
\centerline{
         \epsfxsize=15.0cm
         \epsfbox{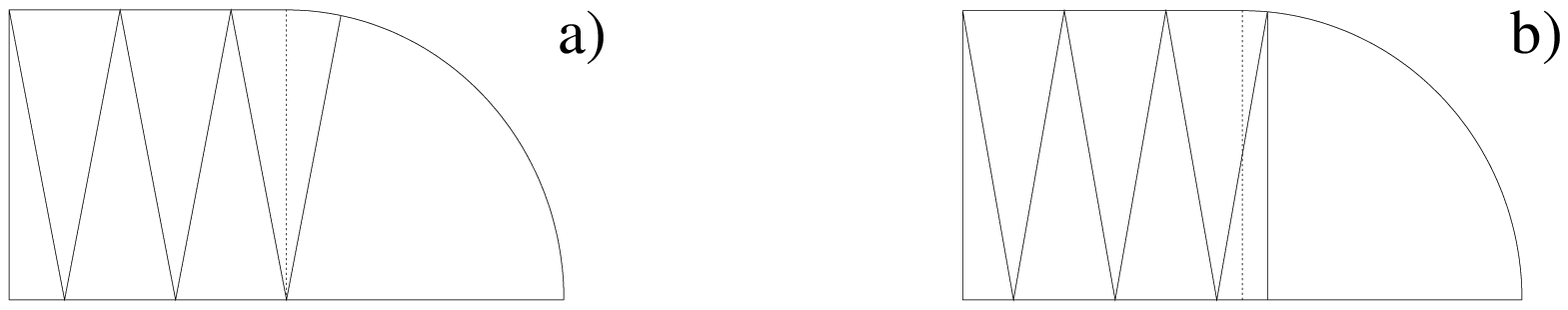}
         }
         \caption[]
            {
 	    Members of the periodic orbit families approaching the
            bouncing ball orbits: a) ($m$, $l$) = (10, 0) and
	    b) ($m$, $l$) = (-11, 1)
	  }
\label{fig:po}
\end{figure}

The bouncing ball limit $|m| \to \infty$ can be reached only for $l = 0$ or 
1, but for all $x$ on the Poincar\'e surface of section. The opposite limit 
$l \to \infty$, (the whispering gallery limit), is possible only for $m = 0$ 
or $1$ and a decreasing $x$ -- interval of starting points. Together, there
are four infinite series of fixed points of the map which are present for all
parameter values $a$. In the $l=0$ case, 
$m$ must be even and positive. The corresponding periodic orbit family 
$(2n,0)$ , see Fig.\ \ref{fig:po}a, has initial conditions
\begin{equation} \label{st0}
x_{(2n,0)} = 0, \quad \phi_{(2n,0)} = \arctan(n/a) 
\qquad \mbox{with} \quad  n=0,1,\ldots
\end{equation}
and 
\begin{equation} \label{L0}
L_{(2n,0)} = \sqrt{(2n)^2 + (2a)^2} + 2, \qquad 
\det({\bf 1}-{\bf M}_{(2n,0)}) = 2 L_{(2n,0)} - 4 .
\end{equation}
Note that $\det({\bf 1}-{\bf M})$ grows linearly with the length $L$ and
thus with the order parameter $n$ which is in contrast to exponential 
growth for strictly hyperbolic dynamics.

The family with $l = 1$ approaching the bouncing ball orbits are formed by 
periodic orbits which start in a corner of the rectangle and  bounce off 
perpendicular to the bottom line in the circle, see Fig.\ \ref{fig:po}b. In 
this case, $m$ must be odd and negative, i.e.\ $m = -2n - 1$. The angle of 
incidence $\theta$ for the bounce in the circle fulfills the condition
\begin{equation}
2 \sin^2\theta + \frac{1}{a} ( 2 n \cos\theta + 1) \sin\theta - 1 = 0,
\quad n = 0,1,\ldots
\end{equation}
with approximate solution
\[ \theta_{n} = \frac{a}{2n + 1} + {\cal O}(n^{-3}). \]
The starting conditions on the Poincar\'e surface are 
\begin{equation}
x_{(-2n-1,1)} = \frac{1}{2 \cos\theta_{n}}, \quad 
\phi_{(-2n-1,1)} = -\frac{\pi}{2} + 2 \theta_{n} < 0. 
\end{equation}
One obtains for the length of the periodic orbits
\begin{eqnarray} \
L_{(-2n-1,1)} &=& 2\sqrt{(\cos\theta_{n} + n)^2 + 
(\sin\theta_{n} + a)^2} + 2\cos\theta_{n}\\
 &=& \sqrt{(2n)^2 + (2a)^2} + 4 - \frac{a^2}{2 n^2} + 
{\cal O}(n^{-3}). \label{L1}
\end{eqnarray}
and for the weight 
\begin{equation}
\det({\bf 1}-{\bf M}_{(-2n-1,1)}) = 
4 \left(\frac{L_{(-2n-1,1)}}{\cos\theta_{n}} 
- 4\right) = 8 n + 5 \frac{a^2}{n} + {\cal O}(n^{-2}),
\end{equation}
(see also appendix \ref{app:A}). Again, the determinant 
increases linearly with the length of the periodic orbits.

An analysis of the periodic orbits $(0,l)$ and $(1,l)$ approaching the 
whispering gallery limit $l\to\infty$ is provided in appendix \ref{app:B}.

\subsection{The Poincar\'e map in the bouncing ball limit} \label{poibb}
The map (\ref{PM}) can be considerably simplified both in the bouncing
ball limit and in the whispering gallery limit. The latter is postponed
to appendix \ref{app:B}. 

For trajectories $(m,0)$, one has to distinguish between
four different cases: 
($m \ge 0$, even), ($m < 0$, even), ($m > 0$, odd) and ($m <0$, odd).
One obtains for ($m \ge 0$, even)
\begin{eqnarray} \label{xbb}
x'_m &=& \frac{\overline{x}}{1 - 2 \overline{x}} 
\left[1 - \frac{4 a^2}{d_m^2} \frac{\overline{x}(1 - \overline{x})^2}
{1 - 2 \overline{x}}\right] + {\cal O}(m^{-4})\\ \label{phibb}
\phi'_m &=& \frac{\pi}{2} - \frac{2 a}{d_m} (1 - 2 \overline{x})
\left[ 1 - \frac{4}{3} \frac{a^2}{d_m^2} 
\frac{1 - \overline{x}(3 - \overline{x}^2)}{1 - 2 \overline{x}}\right] + 
{\cal O}(m^{-5})\\ \nonumber
\mbox{with} & &\\ \nonumber
d_m &=& 2a \tan\phi = m - x + \overline{x}. 
\end{eqnarray}
The length of a trajectory is 
\begin{equation} \label{Lxp}
L(x,\phi) = d + 2 \frac{(1 -\overline{x})^2}{1 - 2 \overline{x}} + 
\frac{2 a^2}{d_m} - 
\frac{4 a^2}{d_m^2} \frac{\overline{x}^2(1 - \overline{x})^2}
{(1 - 2 \overline{x})^2} - \frac{2 a^4}{d_m^3} + {\cal O}(m^{-4}),
\end{equation}
with $\overline{x}(x,\phi) = x + 2a \tan\phi - m > 0$.
Note that $x'$ depends at leading order on $\overline{x}$ only (and not 
independently on both $x$ and $\phi$). For the transfer operator (\ref{BoT}), 
we need the length of a trajectory as function of the initial and final 
points $x$ and $x'$. One obtains 
\begin{eqnarray} \label{Lxx1}
L(x,x') &=& m - x + x' + 2 + \frac{2 a^2}{d_m}\left[1 - \frac{a^2}{d_m^2}
\right] + {\cal O}(m^{-4}) \\ \label{Lxx2}
        &=& \sqrt{d_m^2 + (2 a)^2} + x' - \frac{x'}{1 + 2 x'} + 2 
	+ {\cal O}(m^{-4}) \qquad x,x' \in [0,1]\\ \label{dxx} 
\mbox{with} & & d_m(x,x') = m - x + \frac{x'}{1 + 2 x'} .
\end{eqnarray}
The length of a trajectory after one iteration of the map is thus given by 
its length in the rectangle plus corrections which depend on $x'$ only 
(up to ${\cal O}(m^{-4})$). The mixed second derivatives of $L(x,x')$ showing 
up in the transfer operator (\ref{BoT}) are  
\begin{equation}
\label{dLxx}
\frac{\partial^2 L}{\partial x \partial x'} 
= -\frac{4 a^2}{(d_m^2 + (2a)^2)^{3/2} (1 + 2 x')^2} + {\cal O}(m^{-5}).
\end{equation}

The length spectrum for the Poincar\'e map with $m$ odd or negative is 
obtained by the following replacements in 
(\ref{Lxx1}) -- (\ref{dLxx}):\\[.2cm]
\begin{tabular}{lllll}
($m<0$; even) & $m \to -m$  & $x \to -x$ &  
$x' \to -x'$ & $x\in[0,1];\; x'\in [0,1/3]$\\
($m>0$; odd) & $m\to m+1$ &          &  
$x' \to -x'$ & $x\in[0,1];\; x'\in [0,1/3]$\\
($m<0$; odd) & $m\to -m-1$ & $x \to -x$ & & 
$x\in[0,1];\; x'\in [0,1]$. \\ [.6cm]
\end{tabular}

In a similar way, the approximate map for the $(m,1)$ -- trajectories can be 
constructed. In the bouncing ball limit $m\to\infty$ only 
($m \ge 0$, even) and ($m < 0$, odd) is possible. Again, we
discuss first the case ($m \ge 0$, even), for which we obtain 
\begin{eqnarray} \label{xbb1}
x'_m &=& \frac{\overline{x}}{4 \overline{x} - 1} 
\left[1 + \frac{8 a^2}{d_m^2} \frac{\overline{x}}
{4 \overline{x} - 1} (5\overline{x}^2-4\overline{x}+1) \right] 
+ {\cal O}(m^{-4})\\ \label{phibbi1}
\phi'_m &=& -\frac{\pi}{2} + \frac{2 a}{d_m} (4 \overline{x}-1)
\left[ 1 - \frac{4}{3} \frac{a^2}{d_m^2} 
\frac{2\overline{x}(3 - \overline{x}^2)-1}{4 \overline{x}-1}\right] + 
{\cal O}(m^{-5})\\ \nonumber
\mbox{with} & &\\ \nonumber
d_m &=& m - x + \overline{x} =  2a \tan\phi.
\end{eqnarray}
The length of the trajectory as a function of the initial and final points is  
\begin{eqnarray} \label{Lxx12}
L(x,x')&=&m - x - x' + 4 + \frac{2 a^2}{d_m}\left[1 - \frac{a^2}{d_m^2}\right]
+ {\cal O}(m^{-4}) \\ \label{Lxx22}
        &=& \sqrt{d_m^2 + (2 a)^2} - x' - \frac{x'}{4 x'-1} + 4
        + {\cal O}(m^{-4})\\ \label{dxx2}
\mbox{with}& & d_m(x,x') = m - x + \frac{x'}{4 x'- 1} \qquad
x\in [0,1]; \; x'\in[1/3,1]\\ \label{dLxx2}
\mbox{and}& & \frac{\partial^2 L}{\partial x \partial x'} 
= -\frac{4 a^2}{(d_m^2 + (2a)^2)^{3/2} (4 x'- 1)^2}.
\end{eqnarray}
We recover the ($m<0$, odd) case by replacing  
\[ (m<0; \mbox{odd})\quad  m \to -m-1 \quad  x \to -x
\qquad x\in[0,1];\; x'\in [1/3,1] \]
in (\ref{Lxx12}) -- (\ref{dLxx2}).

\subsection{The transfer operator $\bf T$ in the bouncing ball limit}
\label{sec:T-operbb}
The main advantage of a semiclassical description of quantum mechanics
is to study directly the influence of (parts of) the classical dynamics 
on quantum phenomena. The importance of the near bouncing ball dynamics 
on the quantum spectrum can be understood by analyzing the T--operator 
obtained from the Poincar\'e map in the bouncing ball limit, 
see section \ref{poibb}. 

The $\bf T$ --operator in the bouncing ball limit can be written as
\begin{equation} \label{Tbbl}
T(x,x';k) = T_0(x,x',k) + 
\left\{\begin{array}{ll} 
       T_0(-x,-x',k)& \mbox{if}\quad 0 \le x'\le 1/3\\
       T_1(x,x',k) & \mbox{if} \quad 1/3 < x'\le 1
	\end{array} \right. 
\end{equation}
with
\begin{eqnarray} \nonumber
T_0(x,x') &=&  2 a \sqrt{\frac{k}{2 \pi i}} 
\frac{e^{ik(2 + x' - \frac{x'}{1+2x'}) - i\frac{3}{2}\pi}}{1 + 2 x'}
\sum_{n=0}^{\infty} \left[\frac{e^{i k L_0^-(n)}}{(L_0^-(n))^{3/2}} 
- \frac{e^{i k L_0^+(n)}}{(L_0^+(n))^{3/2}} \right] \\
T_1 (x,x') &=& - 2 a\sqrt{\frac{k}{2 \pi i}} 
\frac{e^{ik(4 - x' - \frac{x'}{4x'-1})-i\frac{3}{2}\pi}}{4 x'- 1}
\sum_{n=0}^{\infty}\left[ \frac{e^{i k L_1^-(n)}}{(L_1^-(n))^{3/2}} 
- \frac{e^{i k L_1^+(n)}}{(L_1^+(n))^{3/2}}\right] \label{T01}
\end{eqnarray}
where $L_{0/1}^{\pm}(n)$ is defined as
\begin{eqnarray}\label{L0+-}
L_0^{\pm}(x,x';n) &=& \sqrt{(2n \pm x + \frac{x'}{1+2x'})^2 + (2 a)^2};\\
L_1^{\pm}(x,x';n) &=& \sqrt{(2n \pm x + \frac{x'}{4x'-1})^2 + (2 a)^2}.
\label{L1+-}
\end{eqnarray}
Here, the lower index corresponds to $l$ = 0 or 1. The upper index $-$ or 
$+$ distinguishes between contributions originating from trajectories with
$m$ even or odd. Note that trajectories in the bouncing ball limit have only 
one caustic in $q$--space inside the circle before returning to the 
Poincar\'e plane both for $l$ = 0 and 1. The additional phases from hard wall 
reflections, see (\ref{nbounce}), have already been incorporated.

\begin{figure}
         \epsfxsize=13.0cm
         \epsfbox{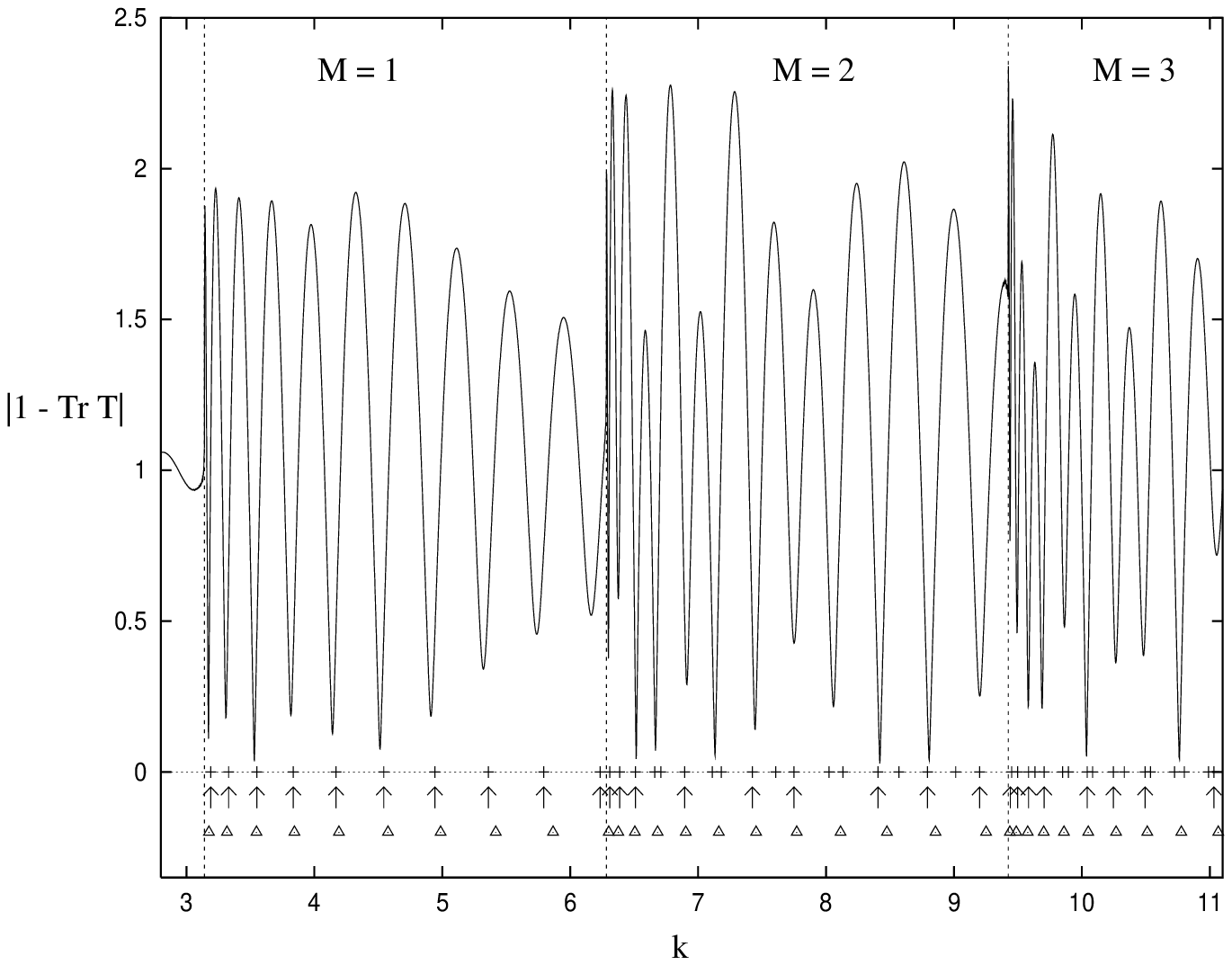}
         \caption[]
            {The absolute value of $1 - \mbox{Tr}{\bf T}$ for $a=5$
             together with the quantum eigenvalues ($+$) and
             the quasi EBK -- solutions ($\triangle$), see Eqn.\ 
	     (\ref{EBK}) in section \ref{sec:EBK}. The bouncing
             ball states are specially marked by arrows.
          }
\label{fig:1-TrT}
\end{figure}

The operator (\ref{Tbbl}) is the main result of this paper. It contains the
dynamics in the stadium in a somewhat counterintuitive way. All 
contributions of trajectories with more than 2 reflections on the circle 
boundary are neglected. In addition, the short orbits for $l$ = 0 or 1 are
represented least accurately. This seems to contradict our common 
understanding
of a semiclassical treatment of quantum mechanics for classically ``chaotic'' 
systems. The quantum spectrum for hard chaos systems is expected to be 
build up collectively by all unstable periodic orbits and the shortest 
periodic orbits are supposed to dominate an expansion of the spectral 
determinants (\ref{detGut}) or (\ref{cumexp}). The stadium billiard posses, 
however, a subset of regular eigenstates, the  so--called bouncing
ball states, which show a nodal pattern very similar to the checkerboard 
pattern obtained for the unperturbed rectangular billiard. It is this
subset of states which can be treated by the approximate
transfer operator (\ref{Tbbl}) alone. This is shown in Fig.\ \ref{fig:1-TrT}.
The leading terms in the cumulant expansion (\ref{cumexp}), i.e.\ 
$\det({\bf 1} - {\bf T}(k)) \approx 1-\mbox{Tr}{\bf T}(k)$, is
plotted here as a function of $k$. The quantum eigenspectrum of the quarter
stadium, marked by crosses on the $k$ axis, is obtained from
the boundary integral method. The eigenvalues having a bouncing ball nodal
pattern are emphasized by arrows. (The bouncing ball states have been 
identified by inspecting individual wave functions). The minima of 
|$1 - \mbox{Tr}{\bf T}(k)$| coincide very well with the eigenvalues 
corresponding
to bouncing ball states found by our subjective criteria. The T--operator
constructed from a Poincar\'e map in the bouncing ball limit fails in other
regions of the spectrum. Some of the states are either completely ignored, 
see e.g.\ at $k\approx 7.6$, or they appear as doublets, where the T -- 
operator expects only a single state, see e.g. around $k\approx 6.7$. The 
latter case 
corresponds to a bouncing ball state interfering with a nearby state 
originating from the non--bouncing ball dynamics. By using higher terms in 
the cumulant expansion (or even the full determinant), non--bouncing ball 
behavior can partly be resolved. A quantization of the full spectrum cannot 
be expected, as important parts of the dynamics have been neglected.
Note, that the trace of $\bf T$ has a cusp at $Re k = m\pi, m=1,2\ldots $.

In the next section, the influence of periodic orbits which appear as 
stationary phase points in the trace of the T--operator will be
studied in more detail. 

\subsection{The trace of $\bf T$ and periodic orbit contributions}
\label{trace}
Let us first concentrate on the trace of the ${\bf T}_0$ -- part in  
the transfer operator (\ref{Tbbl}) which contains contributions from 
trajectories with only one bounce on the circle boundary. The result
for Tr${\bf T}_1$ will be given later. Expanding the length terms
$L^{\pm}_0(n)$ up to ${\cal O}(n^{-2})$ in the exponent and to
leading order in the amplitude, see (\ref{Lxx1}), (\ref{dLxx}), yields
\begin{eqnarray} \label{trbb}
\mbox{Tr}{\bf T}_{0}(k) &=& \sum_{n=0}^{\infty} \mbox{Tr}{\bf T}_{0,n}(k)\\
\nonumber               &=&
\frac{a}{2} \sqrt{\frac{k}{i\pi}}
\sum_{n=0}^{\infty} \frac{1} {n^{3/2}}
e^{i k(\sqrt{(2n)^2 + (2a)^2} + 2) -i\frac{3}{2}\pi}
\\ \nonumber
& &\int_{-1/3}^1 {\rm d}x \frac{1}{1 + 2x}
\left[\exp(ik\frac{a^2 \triangle^-}{2 n^2})-
\exp(2 i k x - ik\frac{a^2\triangle^+}{2 n^2})\right].\\
\nonumber \mbox{with}&&\\
\nonumber \triangle^{\pm}(x) &=& x\pm \frac{x}{1+2x}.
\end{eqnarray}
The phases in front of the trace -- integrals correspond to the length of the 
periodic orbits in the family $(m,l)$ = $(2n,0)$, see (\ref{L0}). The 
negative 
region of integration comes form the second $T_0$ term in (\ref{Tbbl}).

The dominant contribution to each integral is given by the first term 
containing $\triangle^-$ which derives from trajectories with an even number 
of reflections in the rectangle. It is stationary for $x=0$, i.e.\ at the 
starting point of the periodic orbits (\ref{st0}). This becomes obvious after 
the change of variables,
\begin{equation} \label{intp}
\int_{-1/3}^1 {\rm d}x \frac{1}{1 + 2x}
\exp(ik\frac{a^2 \triangle^-}{2 n^2}) =
2 \int_0^{1/\sqrt{3}} dy \frac{1}{\sqrt{1 + y^2}}
\exp(ik\frac{a^2 y^2}{n^2}).
\end{equation}
Approximating the integral straight forward by stationary phase, i.e.\
shifting the limits of integration to infinity, would lead to the standard
periodic orbit amplitudes using (\ref{detM}), here for the periodic orbit
family $(m,l) = (2n,0)$, see (\ref{L0}). The width of the Gaussian is, 
however, increasing with $n$ and the finite limits of integration become 
important for $n\ge a\sqrt{k/\pi}$. The individual contributions to the sum 
(\ref{trbb}) 
thus show a $k$--dependent transition, i.e.\ 
\begin{eqnarray} \label{lbou}
\mbox{Tr}{\bf T}_{0,n}(k) &\to& \frac{1}{2\sqrt{n}}
e^{i k(\sqrt{(2n)^2 + (2a)^2} + 2) -i\frac{3 \pi}{2}}
\mbox{\hspace{1.9cm}}
\mbox{if} \quad n\ll a\sqrt{\frac{k}{\pi}}\\
\label{hbou}
&\to& \frac{a}{2} \sqrt{\frac{k}{\pi}}
\frac{\log 3} {n^{3/2}} e^{i k(\sqrt{(2n)^2 + (2a)^2} + 2) -i\frac{7}{4}\pi}
\qquad \; \mbox{if} \quad n\gg a\sqrt{\frac{k}{\pi}}.
\end{eqnarray}
Contributions from short trajectories in (\ref{lbou}) have the Gutzwiller 
form for isolated unstable periodic orbits with amplitudes 
$|\det({\bf 1}-{\bf M}_n)|^{-1/2} \approx n^{-1/2}$ (see (\ref{L0})) being
independent of $k$. In the other limit $n\gg a\sqrt{\frac{k}{\pi}}$, we 
obtain $e^{i k a^2 y^2/n^2} \approx 1$ within the integration boundaries and 
the stationary 
phase approximation is no longer applicable. \underline{All} trajectories in 
the range of integration give essentially the same contribution to the trace 
as the periodic orbit itself. A whole manifold of orbits build up the
semiclassical weights in the trace in the same manner as manifolds of 
periodic orbits on tori with commensurable winding numbers do in a
semiclassical treatment of integrable systems \cite{Ber77}. In our treatment
as well as in the Berry--Tabor approach \cite{Ber77}, the trace can be 
performed directly and no additional stationary phase approximation is needed 
(in contrast to the Gutzwiller formula). A comparison of (\ref{hbou}) with 
the Berry--Tabor weights obtained for contributions in the rectangle, as 
e.g.\ in Eqn.\ (\ref{inttrap}), unveils the similarity. Note that the 
weights are now explicitly $k$--dependent and decrease like $n^{-3/2}$.
The transition from ``semiclassically integrable'' to ``chaotic'' behavior 
affects mainly the amplitudes. The phase is in both cases essentially given 
by the length of the periodic orbit (\ref{L0}). 

\begin{figure}
         \epsfxsize=13cm
         \epsfbox{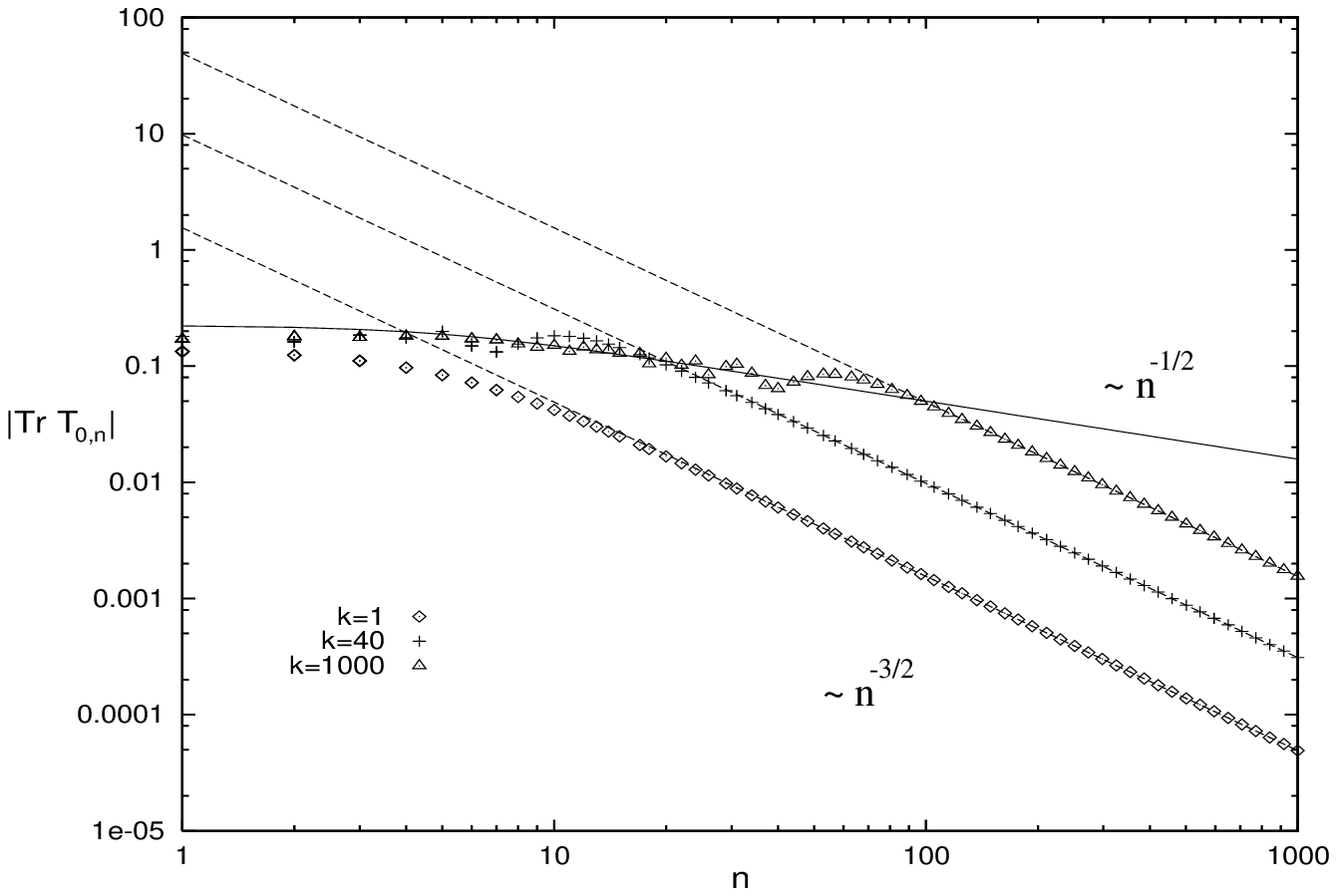}
         \caption[]
            {The modulus of Tr${\bf T}_{0,n}$ as function of the index $n$
	    for different $k$--values showing a $k$--dependent transition 
            from ``chaotic'' to ``integrable'' behavior with increasing $n$; 
	    the full line corresponds to $|{\bf 1} - {\bf M}_n)|$, see 
	    (\ref{L0}), expected for isolated periodic orbits, the dashed 
            lines is the $k$--dependent asymptotic form (\ref{hbou}).
          }
\label{fig:trans}
\end{figure}

The conceptually different treatment for integrable and hard chaos systems 
can thus be rediscovered when studying intermittent dynamics near marginally 
stable boundaries. The contributions to the trace interpolate smoothly 
between the two extremes, the transition region itself is, however, 
$\hbar$ -- (or $k$ --) dependent. To the best of my knowledge, this has been 
shown here for the first time explicitly. The number of terms corresponding 
to contributions of isolated periodic orbits increases like $\sqrt{k}$, the 
turnover occurs for
\begin{equation} \label{thres0}
n_0 \approx \log 3 \; a\sqrt{\frac{k}{\pi}}.
\end{equation}
The transition is indeed rather sharp as can be seen in Fig.\ \ref{fig:trans}.
The modulus of $\mbox{Tr}{\bf T}_{0,n}$ is plotted here versus $n$ for 
different $k$ -- values. The integral (\ref{trbb}) has been calculated 
numerically using the full length formula (\ref{Lxx2}). The oscillations 
occurring in the transition region correspond to the phase change from
$3\pi/2$ to $7\pi/4$ from Eqs.\ (\ref{lbou}) to (\ref{hbou}).

The contributions to the trace from the second term in the integral
(\ref{trbb}) vanish like $1/\sqrt{k}$ for large $k$.  They become, however, 
important for small $k$, especially for the $k$--region below the ground 
state.

In a similar way, the leading contributions to the ${\bf T}_1$ -- operator
in (\ref{Tbbl}) (deriving now from the $L_1^+$ part, see (\ref{L1+-}) and 
thus from trajectories with $m<0$, odd) show a 
transition 
\begin{eqnarray} \label{lbou1}
\mbox{Tr}{\bf T}_{1,n}(k) &\to& \frac{1}{2\sqrt{2 n}}
e^{i k(\sqrt{(2n)^2 + (2a)^2} + 4 -\frac{a^2}{2 n^2})}
\qquad \qquad \quad \; \;\; \mbox{if} \quad n\ll a\sqrt{\frac{k}{\pi}}\\
\label{hbou1}
&\to& \frac{a}{4} \sqrt{\frac{k}{\pi}}
\frac{\log 3} {n^{3/2}} e^{i k(\sqrt{(2n)^2 + (2a)^2} + 4 -\frac{a^2}{2 n^2})
-i\frac{7}{4}\pi}
\qquad \; \mbox{if} \quad n\gg a\sqrt{\frac{k}{\pi}}.
\end{eqnarray}
The phase is again essentially given by the lengths of periodic orbits from 
the family $(m,l)$ = $((-2n-1),1)$, see (\ref{L1}). A $k$--dependent 
transition 
occurs in the amplitude as in (\ref{lbou}), (\ref{hbou}) at a critical
summation index approximately given by the estimate (\ref{thres0}).

Our analysis suggests that a marginally stable boundary as provided here by 
the bouncing ball family is smoothly connected to the outer ergodic regions 
in a semiclassical treatment. We expect that this behavior is generic for 
systems with a mixed phase space structure. Stable islands are surrounded 
by a ``semiclassically stable'' layer and there is a smooth transition 
for semiclassical contributions from both sides of the classically 
disconnected regions. This effect has already been observed by Bohigas et 
al.\ \cite{Boh93,Boh90}. These authors compared the true quantum spectrum in 
the
quartic oscillator with an approximate EBK quantization of a stable island in
this system. They were able to attach quantum states to EBK  results even
beyond the boundary given by the stable islands. This indicates an effective 
enlargement of the stable region into its ``chaotic'' neighborhood by quantum 
effects. Our analysis provides a consistent semiclassical interpretation of 
this phenomenon.

The results so far are based on the particular choice of the Poincar\'e
surface of section. The approach 
presented here is opposite to a semiclassical quantization of the system
on the billiard boundary. The trace of the transfer operator contains then 
only contributions from the shortest periodic orbits, especially 
from the marginal stable family. The Bogomolny method works for this section
as well \cite{Boa94}. Note, however, that corrections to the Gutzwiller 
formula due to intermittency are introduced here through non--classical 
trajectories, when approximating the various traces by stationary phase. 
Note also that a naive summation using the Gutzwiller periodic orbit weights 
(\ref{lbou}) for the stadium would give contributions of the form 
$\sum_{n=1}^{\infty} n^{-1/2} e^{ikL_n}$ which lead to true poles at $k=m\pi$.
It is the $n^{-3/2}$ fall off for large $n$ that prevents the trace 
as well as the spectral determinant from diverging at these points.

One might speculate why the corrections found here have not been observed in 
such a well--studied system like the stadium billiard. Previous studies were 
mainly based on Fourier transformation of the quantum spectrum. The Fourier 
transform exhibits peaks at positions corresponding to the length of periodic 
orbits. The Fourier spectrum is most sensitive to phases of semiclassical 
contributions and only the shortest periodic orbits are resolved when dealing
with smoothed spectra or a finite energy interval. Intermittency as
introduced through the bouncing ball family affects mainly the amplitudes and
contributions of long trajectories. In addition, a $k$--dependence of the
amplitudes as in (\ref{hbou}), (\ref{hbou1}) is washed out by the Fourier
integration. Fourier transformation is thus very insensitive to the influence
of marginally stable behavior. In the next section, I will show that the
bouncing ball spectrum indeed originates from series of orbits approaching
the bouncing ball family and that there is no natural length cut--off for 
trajectories contributing dominantly to the transfer operator.

\subsection{Discrete representation of the transfer operator and quasi
EBK formulas} \label{sec:EBK}
For a further analysis of the operator (\ref{Tbbl}), we proceed in a very 
similar way to Ref.\ \cite{Lau92}. The infinite sums in (\ref{T01}) are 
convergent for $Im(k)\ge 0$ and may be expressed as 
\begin{equation} \label{poisum}
\sum_{n=0}^{\infty} \frac{e^{i k L_0^{\pm}(n)}}{(L_0^{\pm}(n))^{3/2}}
= \frac{1}{2} \frac{e^{i k L_0^{\pm}(0)}}{(L_0^{\pm}(0))^{3/2}} + 
\sum_{m = -\infty}^{+\infty} \int_0^{\infty} {\rm d} n 
\frac{e^{i k L_0^{\pm}(n) -2\pi i m n}}{(L_0^{\pm}(n))^{3/2}}
\end{equation}
using Poisson summation \cite{Tan95}. The integral 
representation (\ref{poisum}) provides an analytic continuation of 
the kernel (\ref{Tbbl}) for $Im(k)<0$ after rotating the axis of
integration ${\rm d}n\to \pm i \, {\rm d}n$ \cite{Tan95,Tan96}. 

For $k$ real, the integrals in (\ref{poisum}) can be evaluated by stationary 
phase.  The prefactors decrease not faster than algebraically and thus vary 
slowly compared to the phase. The stationary phase condition yields
\begin{equation} \label{statphas}
k \frac{\partial L_0^{\pm}(n)}{\partial n} - 2\pi m = 0.
\end{equation}
The solutions of (\ref{statphas}),
\begin{equation}  \label{stsol}
n_m = \frac{1}{2} \left[ \frac{2 a m}{\sqrt{k^2/\pi^2 - m^2}} 
\mp x - \frac{x'}{1+2x'}\right],
\end{equation}
are real for $m \le k/\pi$ only. The saddles for $m<0$ give no 
contribution due to the limits of integration in (\ref{poisum}).
Note, that the sum is not necessarily dominated by short orbits. 
On the contrary, infinitely long trajectories give the main contributions
to the sums in (\ref{T01}) when $k$ approaches $m \pi$ from above.
The singularities at $k = m \pi$ are linked to the cusps appearing in the 
bouncing ball contribution to the integrated trace (\ref{inttr}) when
summing over all repetitions of the marginal stable family.

By skipping the slowly varying first term on the LHS of (\ref{poisum}), 
i.e.\ neglecting again short orbit contributions, one obtains in the 
stationary phase approximation
\begin{eqnarray} \label{Tstph}
T_0(x,x') &\approx&  
- i \frac{e^{2ki(1 + \frac{x'^2}{1+2x'}) + i\frac{\pi}{2}}}{1 + 2 x'}
\sum_{m=1}^{\infty} e^{2\pi i a \sqrt{k^2/\pi^2 - m^2}} 
e^{i \frac{m \pi x'}{1+2x'}} \sin(m\pi x) \\ \nonumber
T_1 (x,x') &\approx& i 
\frac{e^{4ki(1 - \frac{x'^2}{4x'-1})+i\frac{\pi}{2}}}{4 x'- 1}
\sum_{m=1}^{\infty} e^{2\pi i a \sqrt{k^2/\pi^2 - m^2}}
e^{i\frac{m \pi x'}{4x'-1}}\sin(m\pi x).
\end{eqnarray}
The amplitudes are now no longer $k$--dependent.
The sum over $m$ disappears when writing the kernel (\ref{Tbbl}) in the basis 
$\varphi_n = \sqrt{2} \sin(\pi n x), n=1,2,\ldots$, i.e.\ 
\[ T_{m,n} = \int_0^1 dx \int_0^1 dx' 
\varphi_m(x) T(x,x') \varphi_n(x'),\]
which leads to the discrete transfer matrix
\begin{equation} \label{Tnm1}
T_{m,n}(k) \approx 
e^{2\pi i (a \sqrt{k^2/\pi^2 - m^2}+\frac{k}{\pi}+\frac{1}{4})}\; R_{m,n}(k).
\end{equation}
The matrix $\bf R$ is given as
\begin{eqnarray} \label{R-matrix}
R_{m,n}(k) &=&
-i \int_{-1/3}^1 dx' e^{2 k i x'^2/(1+2x')} e^{i m\pi\frac{x'}{1+2x'}}
\frac{\sin(n\pi x')}{1+2x'} \\ \nonumber
&+& i \int_{1/3}^1 dx' e^{2 i k(1-2x'^2/(4x'-1))} e^{i m\pi\frac{x'}{4x'-1}}
\frac{\sin(n\pi x')}{4x'-1}, 
\end{eqnarray}
and is independent of the billiard length $a$. Applying the unitary 
transformation 
\begin{equation}
\tilde{\bf T} = {\bf U}^{-1} {\bf T} {\bf U} 
\qquad \mbox{with} \quad 
U_{m,n} = e^{a \pi i \sqrt{k^2/\pi^2 - n^2}} \delta_{n,m}\; ,
\end{equation}
leads to the more symmetric form  
\begin{equation} \label{Tnm2}
\tilde{T}_{n,m}(k) \approx
e^{2 \pi i \left[\frac{a}{2}(\sqrt{k^2/\pi^2 - n^2} 
+ \sqrt{k^2/\pi^2 - m^2}) +\frac{k}{\pi}+\frac{1}{4}\right]}\; R_{n,m}(k).
\end{equation}

The transfer matrix (\ref{Tnm2}) is now essentially finite with an effective 
dimension
\begin{equation} \label{dimeff}
\mbox{dim}_{\small eff} = \left[k/\pi\right],
\end{equation}
where the brackets $[\;]$ denote the integer part of $k/\pi$. Formula 
(\ref{dimeff}) corresponds exactly to the estimate for the dimension of the 
T--operator given by Bogomolny \cite{Bog92} for the Poincar\'e surface chosen 
here. Neither the dimension nor the matrix $\bf R$ in (\ref{R-matrix}) depend
on the billiard parameter $a$ which enters only through the phase in 
(\ref{Tnm1}), (\ref{Tnm2}). The T--operator derived from the Poincar\'e
map in the bouncing ball limit is expected to reproduce best the bouncing
ball states.  The determinant can be approximated by its leading term in the
cumulant expansion (\ref{cumexp}), i.e.\ $\det({\bf 1} - {\bf T}) \approx
1 - {\rm Tr} {\bf T}$. The phases in front of the matrix $\bf R$ in 
(\ref{Tnm2}) yield a quantization condition for the bouncing ball spectrum.
The matrix $\bf R$ acts as a filter determining the $k$ -- intervals, which
allow for bouncing ball states in principle. A closer analysis shows that
the diagonal elements $R_{mm}$ are dominant and approximately real in the 
region $k \in [m\pi, (m+1)\pi]$. This leads to an EBK--like quantization 
condition 
\begin{eqnarray} \label{EBK}
a \sqrt{\frac{k^2}{\pi^2} - M^2} + \frac{k}{\pi} + \frac{1}{4} = M + N
\qquad \mbox{for }\;  \quad M&=&1,2,\ldots \\ \nonumber
                            N&=&1,2,\ldots, [a\sqrt{2 M + 1} + 5/4].
\end{eqnarray}
$M$ and $N$ act as approximate quantum numbers corresponding to bouncing
ball eigenfunctions with $(M-1)$ nodal lines perpendicular and $(N-1)$ 
nodes parallel to the Poincar\'e surface. (An additional $M$ on the RHS 
of (\ref{EBK}) is introduced for convenience). The cut--off in the 
$N$ -- quantum number originates from the $k$--window given by the $\bf R$ 
matrix. The states with fixed quantum number $M$ are restricted to the 
interval $k \in [M\pi, (M+1)\pi]$ and different $M$ series do not overlap.

In table \ref{tab:EBK}, the bouncing ball eigenvalues (chosen by inspection 
of at the individual wavefunctions) are compared with the quasi EBK 
quantization condition (\ref{EBK}) for $a=5$. The EBK -- solutions are also
marked in Fig.\ \ref{fig:1-TrT}. The importance of the additional terms 
$k/\pi +1/4$ becomes evident when comparing the results with the spectrum of 
the rectangle obtained from to the condition $a \sqrt{k^2/\pi^2 - M^2} 
= N$, i.e.\ $k^2/\pi^2 = M^2 + N^2/a^2$. Again, we find that some bouncing 
ball states expected from the EBK formula (\ref{EBK}) appear to have lost 
their regular nodal pattern due to interference with non--bouncing ball 
states. 

Note, that the T--operator for the rectangle \cite{Lau92} is not recovered in 
the limit 
$a\to\infty$. The $\bf R$ matrix (\ref{R-matrix}) is independent of $a$ and 
the circle boundary can thus not be treated as a small perturbation even for 
large $a$ values. The solutions of (\ref{EBK}) can approximately be written 
in the form 
\begin{eqnarray} \label{ebklim}
\frac{k^2_{M,N}}{\pi^2} &=& M^2 + \frac{1}{a^2}(N - \Delta_{N,M} - 
\frac{1}{4})^2 \\
\nonumber
&\mbox{with}& \quad \Delta_{N,M} = 
\sqrt{M^2 +\left(\frac{N-1/4}{a}\right)^2} 
- M \rightarrow 0 \quad \mbox{for} \quad M \gg \frac{N}{a}.
\end{eqnarray}
A rectangular--like spectrum is achieved in the ``integrable'' limit 
$a\to\infty$, however, in a very special way. First of all, Eqn.\ 
(\ref{ebklim}) contains an extra phase $1/4$ originating from the caustic in 
the circle, which is not present in the rectangle. In addition, the different 
$M$--series always have a finite cut--off for finite $a$, see (\ref{EBK}). 
The number of  states in a given $M$ series increases to infinity only in the 
limit $a\to\infty$.\\

\begin{table}
\begin{center}	
\begin{tabular}{|r|rr|rr|rr|}\hline
   &\multicolumn{1}{c }{$k_{qm}$}
   &\multicolumn{1}{c|}{$k_{ebk}$}
   &\multicolumn{1}{c }{$k_{qm}$}
   &\multicolumn{1}{c|}{$k_{ebk}$}
   &\multicolumn{1}{c }{$k_{qm}$}
   &\multicolumn{1}{c|}{$k_{ebk}$}\\ \hline
 M &\multicolumn{2}{c|}{1}& \multicolumn{2}{c|}{2}& 
\multicolumn{2}{c|}{3}\\\hline \hline
N=1&  3.190 ( 1)&  3.176&  6.309 (11)&  6.301&  9.441 (29)&  9.436\\
  2&  3.329 ( 2)&  3.317&  6.386 (12)&  6.375&  9.497 (31)&  9.487\\
  3&  3.550 ( 3)&  3.547&  6.510 (13)&  6.505&  9.583 (32)&  9.576\\
  4&  3.835 ( 4)&  3.844&            &  6.683&  9.706 (34)&  9.701\\
  5&  4.170 ( 5)&  4.191&  6.896 (16)&  6.904&            &  9.860\\
  6&  4.542 ( 6)&  4.575&            &  7.162& 10.042 (37)& 10.049\\
  7&  4.942 ( 7)&  4.986&  7.423 (19)&  7.452& 10.245 (39)& 10.267\\
  8&  5.361 ( 8)&  5.418&  7.750 (21)&  7.771& 10.494 (41)& 10.511\\
  9&  5.793 ( 9)&  5.865&            &  8.113&            & 10.777\\
 10&  6.233 (10)&  6.325&  8.405 (24)&  8.475& 11.032 (44)& 11.065\\
 11&            &       &  8.793 (26)&  8.855& 11.350 (48)& 11.373\\
 12&            &       &  9.201 (28)&  9.251&            & 11.697\\
 13&            &       &            &       & 11.982 (53)& 12.038\\\hline
\end{tabular}
\caption[soso]{Eigenvalues $k_{qm}$ belonging to bouncing ball eigenstates 
compared with the EBK-like quantization condition (\ref{EBK}) for $a=5$; 
here, $N$, $M$ denote the approximate quantum numbers; the numbers in 
brackets correspond to an enumeration of all states in successive order.}
\label{tab:EBK}
\end{center}	
\end{table}

The determinant $\det({\bf 1} - {\bf T})$ is analytic in each strip
$Re k \in ] m\pi, (m+1)\pi[$, with $m$ integer, but has a cusp at 
$k = m \pi$. 
The non--analytic behavior is introduced through infinitely long trajectories 
contributing in leading order at $k\pi\approx m$, see Eqn.\ (\ref{stsol}). 
These cusps are a consequence of omitting  the bouncing ball contributions, 
which itself exhibit a cusp at integer multiples of $\pi$, see Eqn.\ 
(\ref{inttr}). The full spectral determinant $D(k)$  (\ref{specfak}) is 
analytic and nonanalytic behavior in the bouncing ball contributions is 
canceled by the near--bouncing ball dynamics.

Of special interest is, however, the maximal length of trajectories necessary 
to resolve the regular part of the quantum spectrum at a given wave number 
$k$. Inserting (\ref{ebklim}) in (\ref{stsol}) leads to an estimate for the 
trajectory lengths contributing dominantly to the ground state $(M,N=1)$ in
each $M$ series, i.e.\ to the state located next to the cusp $k = M \pi$. We 
obtain 
\begin{equation} \label{L_est}
L_{max} \approx n_{max} \approx \frac{4}{3} a^2 M \approx \frac{4}{3} a^2 
\frac{k}{\pi}.
\end{equation}
This is in contrast to general semiclassical arguments for bound
systems leading to a cut--off for periodic orbits sums at half 
the Heisenberg time \cite{Ber90}. This transforms for billiard into a
cut--off in the length spectrum according to 
\begin{equation} \label{T_H} 
L_{max} = \pi \overline{d}(k) \approx \frac{A}{4} k ,
\end{equation}
where $\overline{d}(k)$ is the mean level density and $A$ the area of the 
billiard. The estimate (\ref{L_est}) scales differently with the billiard
parameter $a$ and deviates thus especially in the ``integrable'' limit 
$a \to \infty$. 
The cut--off (\ref{T_H}) as derived in Ref.\ \cite{Ber90} is 
based on two main assumptions: the overall validity of the Gutzwiller 
periodic orbit formula and the analyticity of the semiclassical spectral 
determinant in 
a strip containing the real energy axis. Both assumptions which might be
intimately related for bound systems fail here. Note, that this is not 
necessarily true for scattering systems. A semiclassical quantization of 
two examples showing intermittency, the Helium atom \cite{Tan95} and Hydrogen 
in a constant magnetic field for positive energy \cite{Tan96}, could be 
achieved within the Gutzwiller approach. However, the intermittent part of 
the dynamics introduces in these cases as well regular structures in the 
resonance spectrum and non--analyticity in the spectral determinant.

\section{The bouncing ball states}\label{sec:bbstates}
The threshold (\ref{thres0}) can be interpreted as the 
boundary of a region in phase space surrounding the marginal stable bouncing 
ball family. Semiclassical contributions from the dynamics in this region 
have a form similar to the one obtained for stable islands \cite{Boh93,Win94}
or in integrable systems \cite{Ber76,Ber77}. The threshold values 
can be directly translated into momenta in phase space, i.e.\ 
$p^t_x = \pm k \sin\phi_t$ with $\phi_t \approx \arctan(n_{0,1}/a) \approx
\arctan\sqrt{k/\pi}$, see Eqn.\ (\ref{PM}). A {\em semiclassically 
stable island} can thus be defined covering a phase space volume 
$V_{reg} = k^2 \tilde{V}_{reg}$ with  
\begin{eqnarray}\label{vbb}
\tilde{V}_{reg}(k) &=& 4 a \int_{\phi_t}^{\pi/2} {\rm d}\phi
\approx 4 a \left(\frac{\pi}{2} - \arctan\sqrt{\frac{k}{\pi}}\right)\\
\nonumber 
&\approx& 4 a \sqrt{\frac{\pi}{k}} \qquad \mbox{for} \quad k/\pi \gg 1.
\end{eqnarray}
The size of the semiclassical island approaches zero in the limit $k \to 
\infty$ (compared to the volume of the full phase space).
The number of quantum states associated with this island thus increases like 
\begin{equation} \label{nbars}
\overline{N}_{reg}(k) \approx a \left(\frac{k}{\pi}\right)^{3/2} + 
                               {\cal O}(\sqrt{k/\pi}).
\end{equation}
on average. This estimate coincides  with the average increase
of regular states given by the EBK--quantization condition 
(\ref{EBK}). It exceeds previous results by O'Connor and Heller \cite{Con88}, 
who obtained a linear increase in $k$ for the number of localized bouncing 
ball states up to the semiclassical limit $k\to\infty$. 
Note, that a $k^{3/2}$ increase in the number of 
bouncing ball states is still consistent with the Schnirelman  
theorem \cite{Sch74}, as the fraction of regular states compared to
all eigenstates approaches zero in the semiclassical limit. 

\begin{figure}
         \epsfxsize=14cm
         \epsfbox{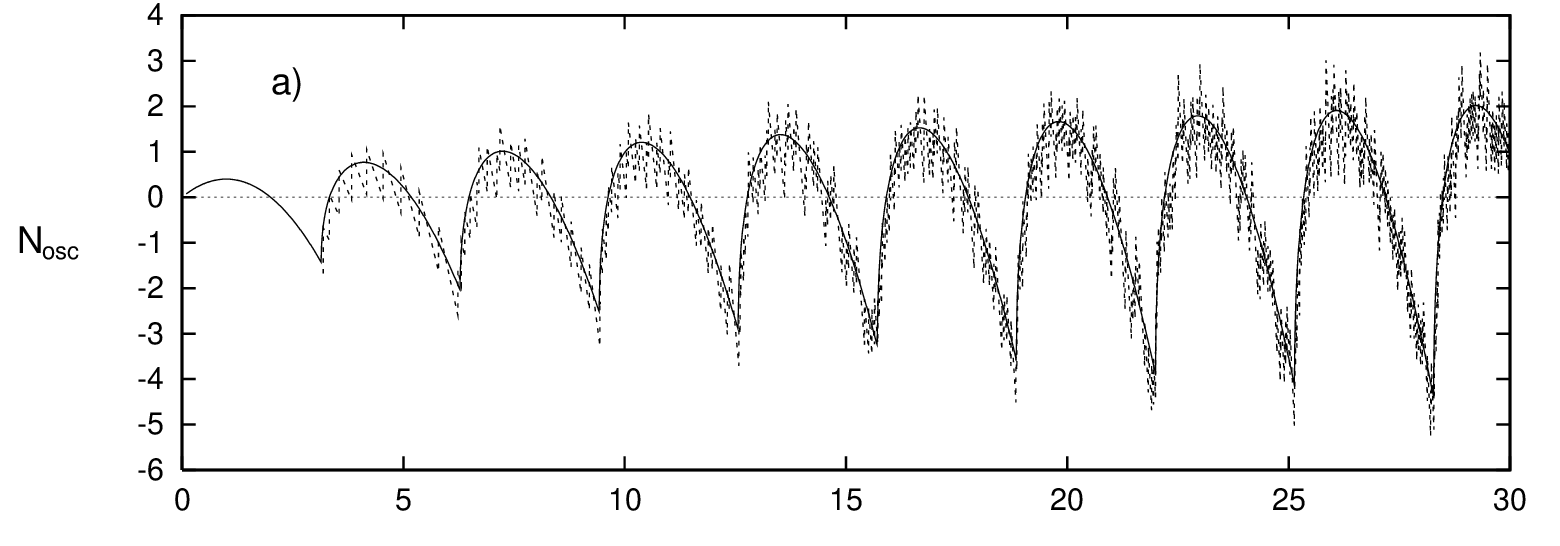}
         \epsfxsize=14cm
         \epsfbox{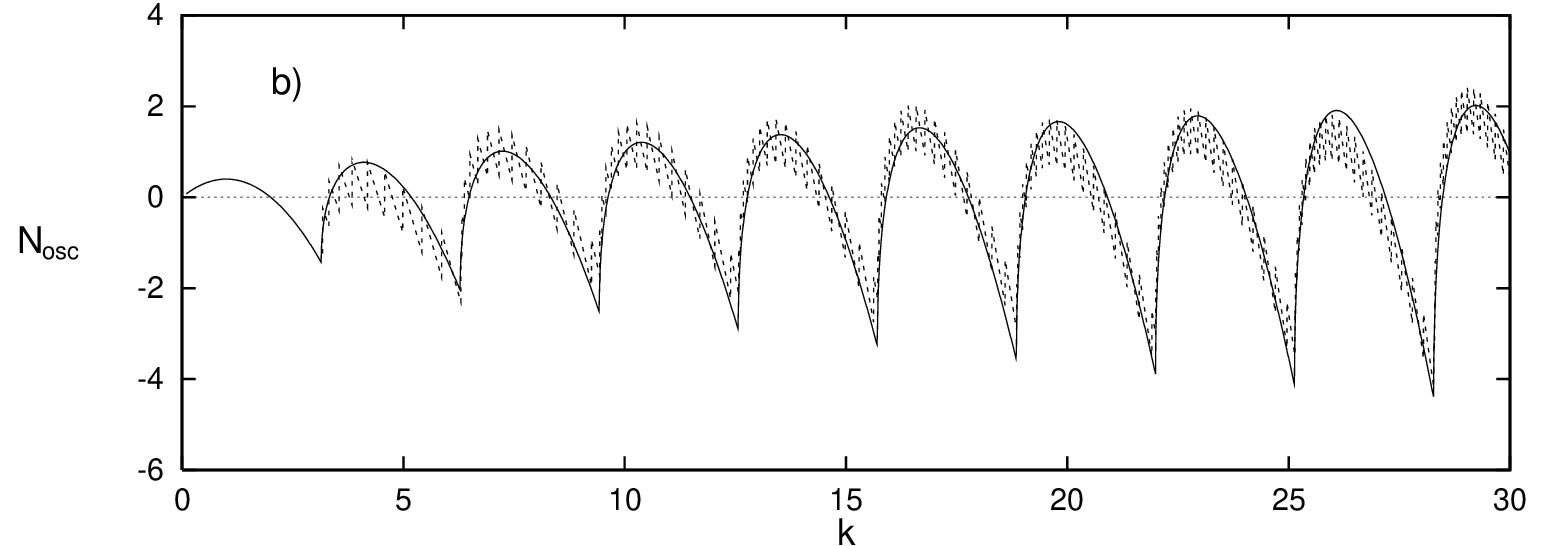}
         \caption[]
            {The oscillating part of the level staircase function (a) for
	     the full spectrum; (b) for the bouncing ball states only.
	     The full line corresponds to the oscillating part of the 
             bouncing ball contributions, see (\ref{inttr}).
	  }
\label{fig:nosc}
\end{figure}

The oscillating part of the level staircase function 
\begin{equation} 
N_{osc}(k) = N(k) - \overline{N}(k)
\end{equation}
shows a strong periodic modulation in the stadium, see Fig.\ \ref{fig:nosc}a. 
($N(k) = \sum_n \theta(k-k_n)$ denotes here the quantum level staircase 
function and $\overline{N}$ its mean part given by the Weyl formula 
\cite{BaH76}).

This modulation coincides with the oscillating part of the bouncing ball 
contributions \cite{Gra92,Sie93,Alo94}, see Eqn.\ (\ref{inttr}). It was shown 
in section \ref{sec:bb}, that the bouncing ball family does not contribute to
individual eigenvalues. From the point of view of
individual states in the spectrum, the oscillatory behavior in 
the level staircase function is caused by a periodic change in the spacings
between neighboring eigenvalues. The spacings are unaffected by 
contributions coming from the classical bouncing ball family.

Taking the concept of a semiclassical stable island seriously, we expect
the spectrum of the stadium billiard to be divided into two different 
subspectra. The majority of states belong to the ``chaotic'' subspace formed 
by non--localized eigenstates (leaving aside the phenomenon of scaring along
short unstable trajectories). Their number increases on average like 
\begin{equation}
\overline{N}_{chaos}(k) = \overline{N}(k) - \overline{N}_{reg}(k) 
\approx \frac{A}{2\pi} k^2 - a \left(\frac{k}{\pi}\right)^{3/2} + {\cal O}(k),
\end{equation}
and the volume of the billiard is $A=a+\pi/4$. The level statistics of 
this subspectrum is expected to follow the GOE prediction of random matrix 
theory. The statistics as well as the average level repulsion is then 
stationary, i.e.\ independent of $k$, in the unfolded 
spectrum. The level staircase function is expected to be 
structureless showing only ``statistical'' fluctuations around the mean value
$\overline{N}_{chaos}(k)$. The regular subspectrum contains the eigenstates 
originating from a quantization of the bouncing ball island (\ref{vbb}). 
The coupling of the dynamics in the ``semiclassically integrable'' region to 
the outer classical motion is weak compared to the mixing in the outer region 
itself. The
level repulsion between bouncing ball states and chaotic eigenstates is
thus small compared to the coupling among non--bouncing ball states itself.
A possible structure in the level staircase function for the regular states
which can be obtained from the EBK-- quantization condition (\ref{EBK}) 
can therefore survive in the full spectrum. This is indeed the case. The 
oscillating part of the EBK -- level staircase function can be defined with 
the help of Eqn.\ (\ref{nbars}), i.e.\  
\begin{equation}
N^{reg}_{osc}(k) = N_{EBK}(k) - \overline{N}_{reg}(k).
\end{equation}
Here, $N_{EBK}(k)$ is given by the number of levels obtained from the
quantization condition (\ref{EBK}) up to a certain $k$ -- value. The result
is shown in Fig.\ \ref{fig:nosc}b, (where the next to leading terms in 
$\overline{N}_{reg}$ have been fitted numerically). The function 
$N^{reg}_{osc}$ shows exactly the same modulations as the full spectrum and
coincides also with the oscillating part of the bouncing ball contributions.
We conclude, that {\em the dominant oscillation in the level staircase
function of the full spectrum is caused by a modulation in the density of 
bouncing ball states only}. 

\section{Conclusion}
I have shown that the marginally stable bouncing ball family in the 
stadium billiard does not contribute to individual quantum eigenvalues. The 
result is confirmed by a semiclassical quantization of the quarter stadium 
using Bogomolny's transfer matrix technique in a representation which 
excludes the bouncing ball orbits explicitly. The regular bouncing ball
quantum states in the spectrum derive semiclassically from the near bouncing 
ball dynamics alone, and these states follow a simple quantization rule. 
The trace of the transfer operator can be approximated by periodic orbit
contributions, which, however, show a $\hbar$--dependent transition from
Gutzwiller to Berry--Tabor--like behavior when approaching the bouncing
ball family. This leads to the concept of a semiclassical island of stability
surrounding the marginal stable family in phase space. The boundary of this 
region is explicitly $\hbar$--dependent and the phase space volume of the
island shrinks to zero (compared to the total volume) in the semiclassical 
limes $\hbar \to 0$ (or $k \to \infty$). The quasi EBK -- quantization 
formula 
can be associated with a quantization of the semiclassical stable island.
The periodic modulations in the level staircase function can be related
to a periodic change in the density of bouncing ball eigenstates. 

The results demonstrate that averaged dynamical properties like ergodicity 
and positive Liapunov exponent are not sufficient to ensure the
applicability of the Gutzwiller trace formula. The ``chaoticity'' of the 
stadium is in a semiclassical sense indeed $\hbar$--dependent, and the 
billiard is for small $k$ closer to an integrable system than to a hard 
chaos one. 

The Gutzwiller periodic orbit weights have to be modified in the whispering 
gallery limit as well, see appendix \ref{app:B}. An accumulation of periodic 
orbits towards a limiting cycle of finite length has also been found in other
systems as e.g.\ in the cardioid billiard \cite{Bru96,Bae96}, in the wedge 
billiard \cite{Han93,Sze94} and in the Anisotropic Kepler Problem 
\cite{Gut90,Tan92}. We expect that a careful treatment as 
outlined in the appendix  will solve problems concerning a semiclassical 
quantization of these systems. \\

The results obtained here for the stadium billiard are expected to be
generic for systems with mixed phase space structure (however complicated
by the existence of island chains and can--tori surrounding the stable island
itself). The stable island influences the classical dynamics in the outer 
chaotic region by creating intermittency.  The regular regions appear to be 
larger than the actual size of the stable island due to the finite phase 
space resolution of quantum mechanics.  The findings explain in a natural 
way the existence of EBK quantum states associated with regions outside a 
stable island as found in \cite{Boh93,Boh90}.
The width of this semiclassically integrable layer is $\hbar$ dependent.  
A semiclassical quantization of stable islands as well as the behavior of 
localized wave functions on classical boundaries and the description of 
tunneling through dynamical separatrices \cite{Boh93,Shu95,Dor95} will be
sensitive to this behavior. The results derived here indicate a failure 
of the Berry--Keating periodic orbit resummation \cite{Ber90} for mixed 
systems due to the intermittency introduced by the stable regions. 

The work presented here is restricted to semiclassical aspects. The influence
of intermittency on Frobenius--Perron and related classical operators 
\cite{Cvi91,Cvi93} is so far best described by a so called 
BER--approximation \cite{Dah94}. The spectra of classical operators are 
directly related to classical \cite{Han84,Cvi96} and semiclassical sum 
rules \cite{Ber85} as well as to spectral statistics \cite{Ber85,And95,Bog96}
 in hard chaos systems. The influence of intermittency 
on these results is still an open issue. \\

\noindent
\underline{Remark:}\\
When completing this article, I became aware of a recent work carried out by
Primack et.\ al.\ \cite{Pri96}. The authors discus diffraction in the Sinai 
billiard by analyzing the Fourier transformation of the true quantum spectrum
in detail. They could indeed relate all deviations from the 
Gutzwiller trace formula to diffraction effects (due to the concave boundaries
in this billiard) \underline{except} for some of the near bouncing ball 
orbits. The results in \cite{Pri96} indicate clearly, that the influence of
intermittency as dicussed in this article can also be seen directly in the 
Fourier spectrum.\\

\noindent
\underline{\bf Acknowledgment}:\\[.2 cm]
I would like to thank Debabrata Biswas, Predrag Cvitanovi\'c, Bertrand 
Georgeot, Kai Hansen, Jon Keating and Mark Oxborrow for stimulating 
discussions and useful comments on the manuscript. The work was supported by 
the Deutsche Forschungsgemeinschaft.

\begin{appendix}
\section{Monodromy matrix in the stadium billiard}
\label{app:A}
The Monodromy matrix used in section \ref{sec:T-oper} describes the 
linearized 
motion near a classical trajectory in phase space in a plane perpendicular 
to the orbit and on the energy manifold. For billiard systems, this plane is 
spanned by the local displacement vectors $\delta q_{\bot}$, 
$\delta p_{\bot}$ pointing perpendicular to the actual momentum of 
the trajectory. An initial displacement is thus propagated according to
\[ 
\left(\begin{array}{c}\delta q^t_{\bot}\\ \delta p^t_{\bot} 
\end{array}\right) = {\bf M}_{q(t),p(t)} 
\left(\begin{array}{c}\delta q^0_{\bot}\\ \delta p^0_{\bot}\end{array} 
\right)\, ,\]
where ${\bf M}$ depends on the path of the underlying trajectory. For
2--dimensional billiards, one obtains 
\[{\bf M}(t) = \left( \begin{array}{cc}
	           1 & L/k\\
	           0 & 1 \end{array} \right)\, ,\]
where $t=L/k$ denotes the time between two bounces at the boundary, $L$ is 
the length of the path and $k = \sqrt{2mE}$. The contribution from a 
reflection at the boundary is 
\[{\bf M}_r = \left(\begin{array}{cc}
                   -1 & 0\\
                \frac{2 k \kappa}{\cos\theta} & -1 \end{array} \right)\, ,\]
where $\kappa$ denotes the local curvature of the boundary, and $\theta$ is 
the angle between the orbit and the normal to the boundary. The $k$ -- 
dependence is scale invariant due to the transformation $\delta q(k) = 
\delta q(k=1)$, $\delta p(k) = k \delta p(k=1)$ and we may set $k=1$ in what 
follows.

Bounces on straight lines in the stadium billiard are treated as free 
flights. The angle of incidence $\theta$ is the same for all successive 
bounces of an orbit in the circle. The length between two of these bounces is
$L = 2 b \cos\theta$ where $b$ denotes the radius of the circle.  
The Monodromy matrix for $l$ free flights in the circle interrupted by $(l+1)$
successive reflections on the circle boundary is thus given by 
\begin{equation} \label{}
{\bf M} = \left(\begin{array}{cc}
                   -(1+2 l) & 2 l b \cos\theta\\
                   \frac{2(l+1)}{b \cos\theta} & -(1+2 l) \end{array} 
\right)\, . 
\end{equation}

\section{The whispering gallery limit}\label{app:B}
The whispering gallery limit of the Poincar\'e map (\ref{PM}) is formed by 
trajectories $(m,l)$ with $m$ fixed and $l\to\infty$ in the notation of 
section \ref{sec:classPM}. These orbits approach the boundary of the billiard 
with an increasing number of bounces in the circle. Only $m = 0$ and $m = 1$ 
is possible in the limit $l\to\infty$. I will consider here the case $m=0$ and
$b = 1$. The approximate whispering gallery map for $m=1$ follows by analogy. 

There is one fixed point for each symbol pair $(m=0,l)$ with starting 
conditions
\begin{equation} \label{powg}
x_l = \cos \left( \frac{\pi}{2}\frac{1}{l+1}\right); \quad \phi_l = 0.
\end{equation}
The angle of incidence $\theta$ for reflections in the circle is
\begin{equation} \theta_l = \frac{l}{l+1} \frac{\pi}{2}.\end{equation}
The length of the corresponding periodic orbit approaches a constant, i.e.\
\begin{equation} 
L_l = 2 a + 2(l+1) \cos\theta_l = 2 a + \pi - 
\frac{\pi^3}{24} \frac{1}{(l+1)^2} + {\cal O}(l^{-4}).
\end{equation}
The Monodromy matrix along the orbits starting on the Poincar\'e surface 
of section can be deduced from appendix \ref{app:A} and is
\begin{equation} \label{Mwg}
{\bf M}_l = \left(\begin{array}{cc}
           1 & 2 a \\
           \frac{2(l+1)}{ \cos\theta_l} & \frac{4a(l+1)}{\cos\theta_l} + 1
                  \end{array} 
\right)\, . 
\end{equation}
The periodic orbit weighting factor $|\det({\bf 1}-{\bf M}_l)|$ is thus 
\begin{equation}
|\det({\bf 1}-{\bf M}_l)| = 4 a \frac{l+1}{\cos\theta_l} = 
\frac{8a}{\pi} (l+1)^2 + \frac{\pi}{3} a + {\cal O}(l^{-2}).
\end{equation}
The determinant and thus the largest eigenvalue $\Lambda_l$ of the Monodromy 
matrix increase quadratically with the symbol index. The main difference 
compared to the bouncing ball limit is the convergence of the period $L_l$ 
towards a finite value: the length of the billiard boundary. The whispering 
gallery limit therefore introduces no intermittency. The Liapunov -- exponent 
of periodic orbits,
\[ \lambda_l = \frac{\log\Lambda_l}{L_l} \sim \log l, \]
diverges logarithmically and the boundary orbit $(0,l=\infty)$ is infinitely
unstable.  

I will show below that the isolated orbit condition is again violated when
taking the trace of the transfer operator. A naive summation of Gutzwiller 
periodic orbit weights would overestimate the whispering gallery 
contributions 
considerably. The breakdown of the stationary phase condition is caused here 
by the accumulation of periodic orbits, i.e.\ stationary phase points,
near the billiard boundary.\\

The phase space area of starting conditions on the Poincar\'e surface for 
trajectories with the same $l$--value is centered on the corresponding 
periodic orbit. Its size shrinks to zero
in the limit $l\to\infty$ both in the $x$ and $\phi$ coordinate.
The transfer operator in the whispering gallery limit is thus directly
given by the Jacobian matrix (\ref{Mwg}) in local coordinates
$\delta x = x - x_l$, $\delta\phi = \phi - \phi_l$ centered on the fixed 
points (\ref{powg}). The orbit length is approximately given by 
the quadratic form
\[ L_l(\delta x, \delta x') = L_l + \frac{1}{2} 
\left(\begin{array}{c}\delta x\\ \delta x' \end{array}\right)^T 
{\bf J}_l
\left(\begin{array}{c}\delta x\\ \delta x' \end{array}\right)\]
with 
\[{\bf J}_l = \left( \begin{array} {cc} 
                         \frac{\partial^2L}{\partial x \partial x}&
                         \frac{\partial^2L}{\partial x \partial x'}\\
                         \frac{\partial^2L}{\partial x' \partial x}&
                         \frac{\partial^2L}{\partial x'\partial x'}
       \end{array} \right)_{x=x'=x_l}
 = \left( \begin{array} {cc}   \frac{1}{2a}& -\frac{1}{2a}\\ 
	                -\frac{1}{2a}& 
                         \frac{1}{2a} + \frac{2(l+1)}{\cos\theta_l}
       \end{array} \right) \, ,
\]
where $L_l$ denotes the length of the periodic orbit.  When taking the trace 
of $\bf T$ we are interested in trajectories satisfying 
$\delta x = \delta x'$ which demands $\delta \phi = 0$. More important is, 
however, the restrictions on the $\delta x$ -- interval, i.e.\
\begin{equation} \label{bwg}
-\frac{\pi}{2} \frac{\cos\theta_l}{(l+1)(2l+1)} 
\le \delta x < 
\frac{\pi}{2} \frac{\cos\theta_l}{(2l+3)(2l+1)}. 
\end{equation}
The interval length and thus the integration range decreases like $l^{-3}$.

The whispering gallery contribution to the trace of the 
$\bf T$ -- operator are
\begin{equation}
\mbox{Tr}{\bf T}_{wg} = 
\sum_{l=0}^{\infty} \frac{1}{\sqrt{2\pi i}} \sqrt{\frac{k}{2a}} e^{i k L_l - 
i \frac{3}{2}(l+1) \pi} \;
2 \int_0^{\overline{\delta}} {\rm d}x\, e^{ik \frac{l+1}{\cos\theta_l}x^2},
\end{equation}
where the upper and lower limits in (\ref{bwg}) have been approximated by 
their mean value 
\[\overline{\delta} = \pi \frac{\cos\theta_l}{(2l+1)(2l+3)} \approx 
\frac{\pi^2}{2} \frac{1}{(l+1)(2l+1)(2l+3)}.\]
The interval length decreases faster than (the square root
of) the prefactor in the exponent for increasing $l$. A stationary phase 
approximation is not justified for $l\to\infty$, and we obtain (as in the
bouncing ball limit) a transition behavior
\begin{eqnarray}
\mbox{Tr}{\bf T}^l_{wg} &\to& \frac{1}{\sqrt{|\det({\bf 1}-{\bf M_l}|} }
e^{i k L_l - i \frac{3}{2}(l+1) \pi} \qquad
\qquad \mbox{for} \quad l\ll \left(\frac{\pi}{2} k\right)^{1/4}\\
\mbox{Tr}{\bf T}^l_{wg} &\to& \sqrt{\frac{\pi k}{a}} 
\frac{\cos\theta_l}{(2l+1)(2l+3)} 
e^{i k L_l - i \frac{3}{2}(l+1) \pi -i\frac{\pi}{4}} 
\qquad \mbox{for} \quad l\gg \left(\frac{\pi}{2} k\right)^{1/4}.
\end{eqnarray}
Note, that the number of periodic orbits which can be treated as isolated
stationary phase points,
increases only at a rate proportional to $k^{-1/4}$. The contributions for
large $l$ fall off like $l^{-3}$ and thus faster than in the bouncing 
ball case. In addition, there is a cancelation between the $(0,l)$ and the 
$(1,l)$ family which show the same overall behavior but appear with opposite 
signs. The whispering gallery contributions are thus small compared to
near bouncing ball contributions, at least for $a\ge 1$.
\end{appendix}

\end{document}